\documentclass[manuscript]{acmart}

\usepackage{multirow}%
\usepackage{amssymb,amsfonts}%
\usepackage{amsthm}%
\usepackage[title]{appendix}%
\usepackage{textcomp}%
\usepackage{algorithm}%
\usepackage{listings}%
\usepackage{subcaption}%
\usepackage{cleveref}%
\usepackage{enumitem}%

\newcommand{\proposedcove}{{\sc CoVE}}
\newcommand{\hgruforrec}{{\sc HGRU4Rec}}
\newcommand{\gruforrec}{{\sc GRU4Rec}}

\newcommand{\sasrec}{{\sc SASRec}}

\newcommand{\bpr}{{\sc BPR}}
\newcommand{\fpmc}{{\sc FPMC}}
\newcommand{\pfive}{{\sc OpenP5}}

\newlist{Properties}{enumerate}{2}
\setlist[Properties]{label=Property \arabic*,itemindent=*}

\AtBeginDocument{%
  }

\begin{document}

\title{Compositions of
  Variant Experts for Integrating Short-Term and Long-Term Preferences}


\author{Jaime Hieu Do}
\email{dinhhieu.do.2020@smu.edu.sg}
\orcid{0000-0003-3923-4499}
\affiliation{%
  \institution{Singapore Management University}
  \country{Singapore}
}
\author{Trung-Hoang Le}
\email{lthoang@hcmus.edu.vn}
\orcid{0000-0002-2349-482X}
\affiliation{%
  \institution{Faculty of Information Technology, University of Science, Ho Chi Minh City}
  \country{Vietnam}
}
\affiliation{%
  \institution{Vietnam National University, Ho Chi Minh City}
  \country{Vietnam}
}
\author{Hady W. Lauw}
\email{hadywlauw@smu.edu.sg}
\orcid{0000-0002-8245-8677}
\affiliation{%
  \institution{Singapore Management University}
  \country{Singapore}
}

\renewcommand{\shortauthors}{Do et al.}

\begin{abstract}
  In the online digital realm, recommendation systems are ubiquitous and play a crucial role in enhancing user experience. These systems leverage user preferences to provide personalized recommendations, thereby helping users navigate through the paradox of choice.
  This work focuses on personalized sequential recommendation, where the system considers not only a user's immediate, evolving session context, but also their cumulative historical behavior to provide highly relevant and timely recommendations.
  Through an empirical study conducted on diverse real-world datasets, we have observed and quantified the existence and impact of both short-term (immediate and transient) and long-term (enduring and stable) preferences on users' historical interactions.
  Building on these insights, we propose a framework that combines short- and long-term preferences to enhance recommendation performance, namely Compositions of Variant Experts (\proposedcove). This novel framework dynamically integrates short- and long-term preferences through the use of different specialized recommendation models (i.e., experts). Extensive experiments showcase the effectiveness of the proposed methods and ablation studies further investigate the impact of variant expert types.
\end{abstract}


\maketitle

\section{Introduction}

Recommendation systems have become an integral part of the online digital realm, significantly enhancing user experience by providing personalized suggestions. Leveraging user preferences from past interactions, these systems help users navigate through the vast array of available choices faster by providing relevant recommendation.
Modern recommendation systems, particularly sequential recommendation, consider the user's temporal preferences, including recent interactions such as clicks, views, purchases, and location check-ins, to provide recommendations.

\begin{figure*}[t]
  \centering
  \includegraphics[width=1\textwidth]{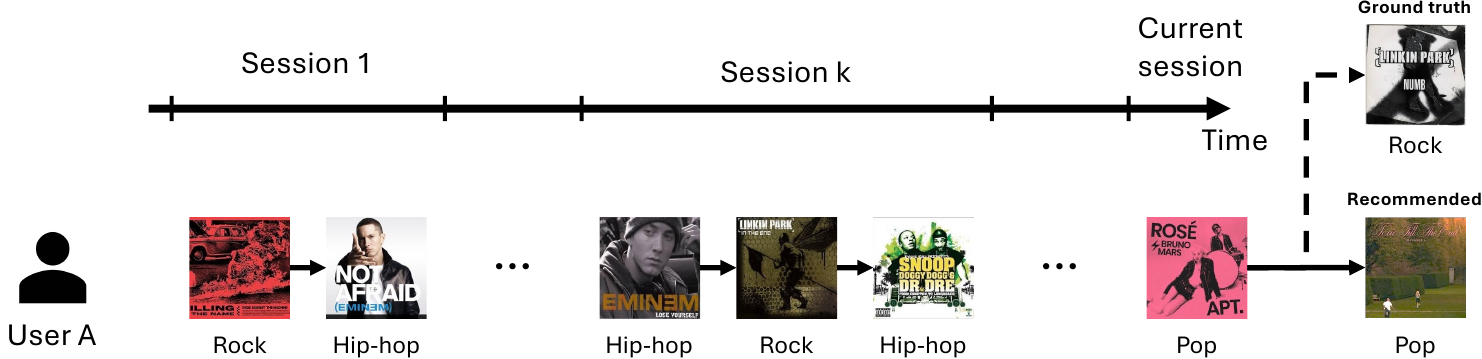}
  \caption{User A might want to be recommended a rock song based on their long-term interest in rock and hiphop, rather than a pop song based on their short-term interest in a trending pop song.}%
  \label{fig:motivating-example}%
  \Description{}
\end{figure*}

There is a distinction between \emph{session-based} and \emph{session-aware} recommendation systems. Session-based recommenders focus on capturing sequential user behavior within a single session, reflecting a user's short-term preference and aiming to predict the next item in that sequence. Conversely, session-aware recommenders incorporate both the current session context and the user's historical behavior, thereby encompassing their long-term preference. This may enable session-aware recommendation systems to offer more accurate personalized recommendations.

A critical aspect of session-aware recommendation systems is the differentiation between short- and long-term user preferences. The former are derived from the user's recent interactions within the current session, while the latter are based on historical behavior over a more extended period. An illustrative example is shown in ~\autoref{fig:motivating-example}, user $A$ is on a streaming music platform. Her historical sessions show a strong preference for rock and hip-hop, with representative artists such as {\em ``Rage Against the Machine''}, {\em``Eminem''}, and {\em``Linkin Park''}. However, in the current session, however, she is listening to  {\em``APT.''} by {\em``Bruno Mars and Rosé''}, a trending pop song, indicating a short-term preference for pop. The challenge is deciding whether to recommend more pop songs or stick with user $A$'s usual hip-hop/rock. If only short-term preferences are considered, the system may recommend more pop songs, which may not align with her typical tastes. Conversely, focusing solely on long-term preferences may overlook her current interest. Understanding and effectively exploiting these preferences can lead to more accurate and personalized recommendations.

Integrating short-term and long-term preferences poses several challenges. Existing session-based recommendation models often focus on capturing short-term preferences, neglecting long-term preferences. On the other hand, traditional recommendation models that capture long-term preferences may not be suitable for session-aware recommendations due to the lack of ability to model sequential user behavior.  To address this gap, we propose compositions of variant experts (\proposedcove), a novel framework that integrates both short- and long-term preferences to enhance recommendation performance by leveraging the strength of multiple types of experts working in concert.



  {\bf Contribution.} First, we explore the existence and impact of short- versus long-term preferences in session-aware recommendation systems in~\autoref{sec:short-vs-long-term-preferences}. Second, we propose the Compositions of Variant Experts (\proposedcove) with two main variants, which dynamically integrate short- and long-term preferences in~\autoref{sec:method}. Third, in~\autoref{sec:exp}, we conduct extensive experiments to demonstrate the effectiveness of the proposed models and analyze the impact of the variant expert types, the number of experts, as well as the gating mechanism.



\section{Short-Term and Long-Term Preferences in Session-Aware Recommendation}
\label{sec:short-vs-long-term-preferences}



\begin{table}
  \centering
  \caption{Notations}
  \begin{tabular}{c|l}
    \toprule
    Symbol                                                                & Description                                                               \\
    \midrule
    $\mathcal{P}=\{1, \dots, p, \dots, N\}$                               & the set of $N$ items, $N=|\mathcal{P}|$                                   \\
    $\mathcal{U}=\{1, \dots, u, \dots, M\}$                               & the set of $M$ users, $M=|\mathcal{U}|$                                   \\
    $S_t^u = \left\langle p_{t,1}, p_{t,2}, \dots, p_{t,K} \right\rangle$ & user $u$'s session at time period $t$ with a total of $K$ items           \\
    $C^u_{1:T} = \left\langle S_1^u, S_2^u, \dots, S_{T}^u \right\rangle$ & set of user $u$'s sessions from time period $1$ up to time period $T$     \\
    $\left\{ f_1, f_i, \dots, f_n \right\}$                               & the set of $n$ experts                                                    \\
    $f\left(u, C^u_{1:T}\right)\rightarrow \mathbb{R}^N$                  & predicted scores of $N$ items                                             \\
    $h_{t}^u$                                                             & user $u$ representation up to time $t$                                    \\
    $\Psi(p)$                                                             & item $p$ representation                                                   \\
    $\beta(p)$                                                            & item $p$ bias                                                             \\
    $g(p_{m,t})$                                                          & $n-$dimensional output of the gating function w.r.t. item input $p_{m,t}$ \\
    $g(p_{m,t})_i$                                                        & $i-$th component of $g(p_{m,t})$                                          \\
    \bottomrule
  \end{tabular}
\end{table}

Let $\mathcal{P}$ be the universal set of items, where $N=|\mathcal{P}|$ is the total number of items. Let $\mathcal{U}$ be the set of users, $M=|\mathcal{U}|$ is the total number of users.
A user $u \in \mathcal{U}$ interacts with several items within a short span of time, denoted $t$, forming a \textit{session} $S^{u}_t = \left\langle p_{t,1}, p_{t,2}, \dots, p_{t,K} \right\rangle$, where $p_{t,k}\in\mathcal{P}, \forall k\in\{1,2,\dots,K\}$, $K$ is the total number of items in session $S_t^u$.
All of these user interactions from multiple sessions in chronological order $C_{1:T}^u = \left\langle S_1^u, S_2^u, \dots, S_{T}^u \right\rangle$ reflect both user \textit{short-term} within a session and \textit{long-term} preferences which are items across all the sessions. We presuppose the most recent session $S_T^u$ of $C_{1:T}^u$ is the current browsing session of user $u$. The main task is to predict the next probable item $p_{T,K+1}$, where $K$ is the total number of items in the current session $S_{T}^u$.

Let $f$ be an ``expert'', in this case, a model, defined as a function $f\left(u, C^u_{1:T}\right)\rightarrow \mathbb{R}^N$ that maps an input set of sessions from user $u$, $C^u_{1:T}$, to an $N-$dimensional vector of real values, where each value denotes the predicted ranking score for an item. In this article, we use the terms ``expert'' and ``preference model'' interchangeably.

Here, we discuss the categorization of $f$ into two, i.e., short-term expert and long-term expert. The following properties are useful in determining to which category an expert belongs.

\begin{Properties}
  \item ({\sc Single Session Property}) \label{prop:single} The model processes only one session at a time, without any cross-session interactions. Specifically, when provided with multiple sessions, the model considers only the most recent one, i.e., $f\left(u, C^u_{1:T}\right) = f(u, S_T^u)$.
  \item ({\sc User Independence Property}) \label{prop:user-independence} Given an identical sequence of input items, the model, denoted as $f$, always produces the same output regardless of user-specific information. Formally, for any two users $u \neq u'$, $f(u, C_{i:j}^u)=f(u', C_{i':j'}^{u'}), \forall C_{i:j}^u = C_{i':j'}^{u'}, i\leq j, i' \leq j'$.
\end{Properties}

Based on these two properties, short-term and long-term preference model are formally defined as follows:

\begin{definition} ({\sc Short-term preference model}) \label{def:short-term}
  A model is considered as a short-term preference model if both \ref{prop:single} and \ref{prop:user-independence} hold.
  In particular, a short-term preference model $f\left(u, C_{1:T}^u \right)$ takes as input a single sequence of items $S_T^u$ (following \ref{prop:single})
  and aims to predict the next probable item $p_{T,K+1}$. The model is user independent, meaning for the same input sequence, it always produces the same output regardless of the user, i.e., $f\left(u, S_{t}^u\right) = f(u', S_{t'}^{u'}),\forall S_{t}^u=S_{t'}^{u'}, u\neq u'$.
\end{definition}

\begin{definition} ({\sc Long-term preference model})
  \label{def:long-term}
  A model is considered as a long-term preference model if either \ref{prop:single} or \ref{prop:user-independence} does not hold.
  A long-term preference model $f \left( u, C_{1:T}^u \right)$ considers historical sessions $C_{1:T}^u$ or user $u$, or both as inputs to predict the most probable item $p_{T,K+1}$. The long-term model is user-dependent, meaning for two users $u$ and $u'$, $u\neq u'$, the model may produce different outputs given the same historical sessions, i.e., $\exists f(u,C)=f(u', C)$. Additionally, a model that considers more than one session is also a long-term preference model.
  The long-term preference model may learn user underlying preferences, concentrated into a dense representation $h^u$ or model the inter-session relations instead of one single session.
\end{definition}

\begin{figure}[t]
  \centering
  \includegraphics[width=0.8\textwidth]{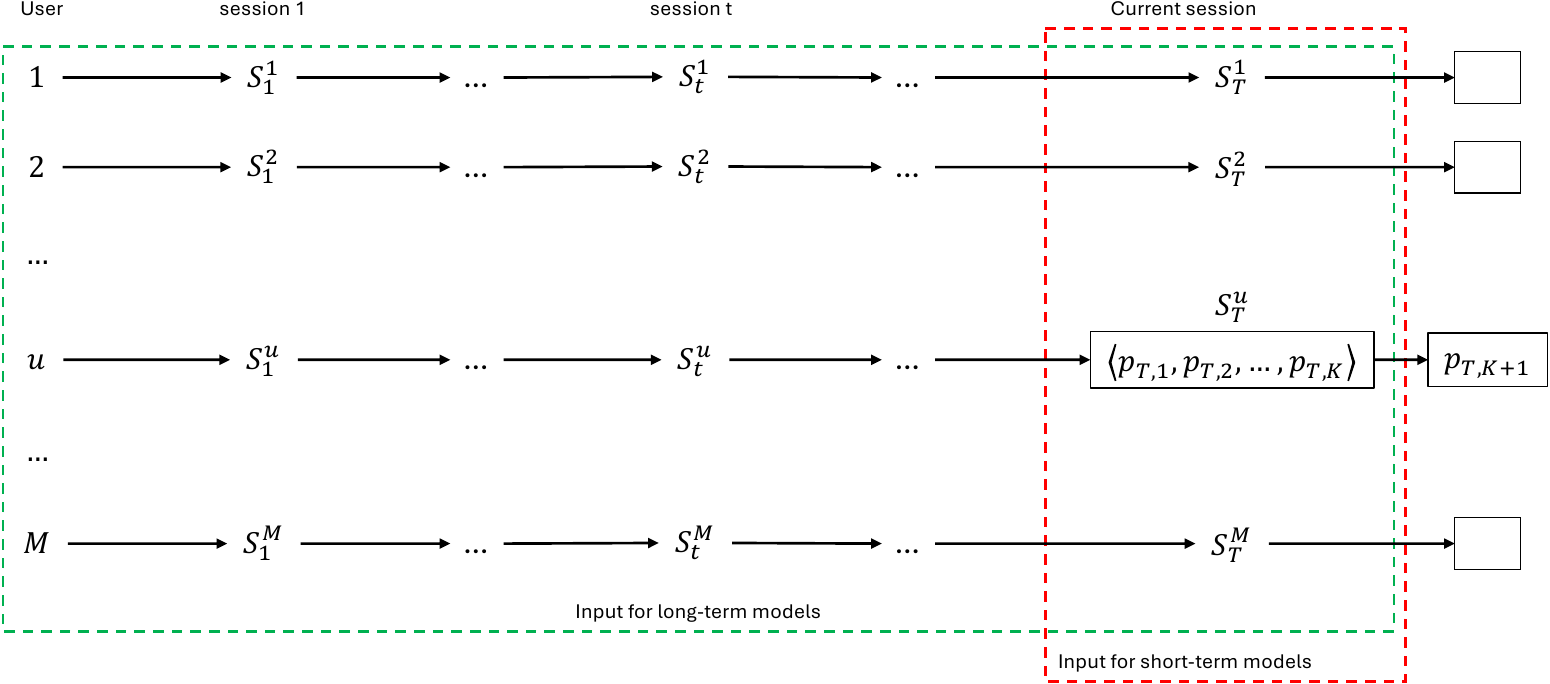}
  \caption{An illustrative example distinguishing inputs for long-term and short-term preference models.}%
  \label{fig:long-term-versus-short-term-preference-inputs}%
\end{figure}

A short-term preference model operates solely on the interactions occurring within the current session, without incorporating any external information such as user demographics, historical preferences, or side data. This makes it well-suited for scenarios where long-term user data is unavailable or irrelevant.
A notable example of a short-term preference model is \gruforrec~\cite{gru4rec1}, which utilizes Gated Recurrent Unit (GRU)~\cite{chung2014empirical}, widely used to model sequential data such as text. \gruforrec\ processes a sequence of recently interacted items to generate recommendations, focusing exclusively on short-term engagement patterns rather than long-term user history. For better illustration, we highlight the inputs of short-term preference model in the red box in~\autoref{fig:long-term-versus-short-term-preference-inputs}.


A long-term preference model leverages stored user information--such as user representation or historical interactions--to enhance future recommendation predictions. It could vary from general collaborative filtering to sequential models which utilize previous sessions together with the current one. For instance, Bayesian Personalized Ranking (\bpr)~\cite{rendle2009bpr} takes a user as input to produce a personalized list of recommended items (green box in~\autoref{fig:long-term-versus-short-term-preference-inputs}).
Another example instance of long-term preference model, \hgruforrec~\cite{quadrana2017personalizing}, uses an additional GRU layer to model all previous sessions of the same user into a condensed representation, which is combined with the current session for next-item prediction.



To further verify the existence of short-term and long-term preferences in session-aware recommendation systems, we conduct an empirical analysis on the public Diginetica\footnote{\url{https://competitions.codalab.org/competitions/11161}} and RetailRocket\footnote{\url{https://www.kaggle.com/datasets/retailrocket/ecommerce-dataset}} datasets. Here, we particularly analyze the recommendation performance of two candidate models: one that only models long-term preferences without short-term preferences, and the other vice versa. In particular, we use two representative models, \gruforrec~\cite{gru4rec1}, which models short-term preferences through sequential session interactions, and \bpr~\cite{rendle2009bpr}, which captures long-term user preferences through user-item interactions. More details on the datasets are provided in~\autoref{subsec:experiment-setup}.
In these datasets, user interactions occur in sessions, where each session consists of a sequence of item interactions within a given time window, and users may have multiple sessions over time.
We hypothesize that there exist both short-term and long-term preferences in session-aware recommendations. To validate this hypothesis, we evaluate the recommendation effectiveness of \bpr\ and \gruforrec\ using the same training data but formatted to suit each model’s input requirements. \bpr\ processes a user-item interaction matrix without sequential information, while \gruforrec\ takes a sequence of recently interacted items as input. Both models generate ranked recommendation lists as next item prediction.
To determine whether an item aligns more with short-term or long-term preferences, we compare the ranking of the ground truth item in the test set. If \bpr\ ranks the item higher than \gruforrec, we classify it as a long-term preference (bit = 1). Conversely, if \gruforrec\ ranks it higher, we classify it as a short-term preference (bit = 0). In cases where both models rank the item equally, we assign a value of 0.5.


\begin{figure*}[t]
  \centering
  \hfill
  \begin{subfigure}[]{0.4\textwidth}
    \centering
    \includegraphics[width=\textwidth]{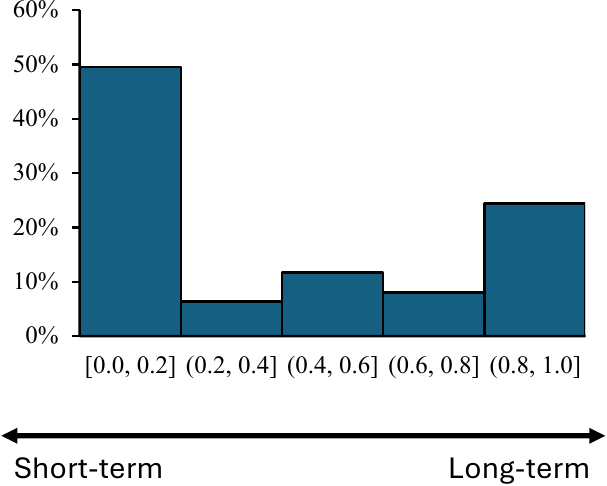}
    \caption{Diginetica}
    \label{subfig:preference-histogram-diginetica}
  \end{subfigure}
  \hfill
  \begin{subfigure}[]{0.4\textwidth}
    \centering
    \includegraphics[width=\textwidth]{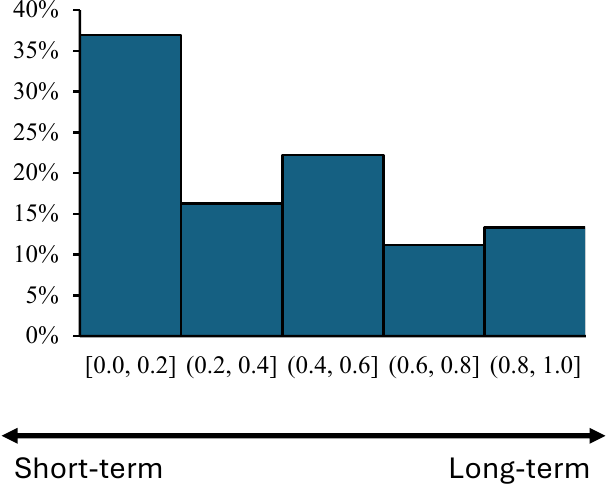}
    \caption{RetailRocket}
    \label{subfig:preference-histogram-retailrocket}
  \end{subfigure}
  \hfill\mbox{}
  \caption{Histogram of user preference values}%
  \label{fig:preference-values}%
\end{figure*}

The histograms in Figure~\ref{fig:preference-values} visualize the distribution of long-term bits (as a fraction) across the test datasets.
While some users have extreme long-term preferences, with an average long-term bit value of $1$, others do not have any long-term interest with a value of $0$. The average values are $0.39$ and $0.38$ for Diginetica and RetailRocket datasets respectively, indicating that short-term preferences are slightly more dominant.

Through this empirical analysis, we observe that there exists both short-term and long-term preferences in personalized sequential recommendation, which can drive user behavior differently given their personal preferences. Effectively modeling and balancing these factors holds promise for enhancing recommendation quality, thereby improving user engagement.

\section{Compositions of Variant Experts (\proposedcove)}\label{sec:method}
In this section, we discuss the use of variant experts for session-aware recommendations. The core idea is to leverage different types of both short-term and long-term preference models as experts, each providing unique insights into the data, that collectively combine the strengths of different experts to create a more effective recommendation model.


\subsection{Preliminary}
Here, we discuss background knowledge of well-known Mixture-of-Experts methods including continuous mixture of experts and sparse mixture of experts.

  {\bf Mixture-of-Experts.} (MoE) model~\cite{jacobs1991adaptive} is designed to increase the model's capacity while maintaining low computational costs by using multiple ``experts''.
This gating mechanism allocates input tokens to different experts, making each expert specified with distinct input data.
There are several ways to implement the gating network $g(x)$~\cite{clark2022unified,hazimeh2021dselect,zhou2022mixture}, but a simple and commonly used approach is to apply a softmax function over the logits produced by a linear layer~\cite{shazeer2017outrageously}:
\begin{equation}
  g(x):=\text{Softmax}(x \cdot W_g)
\end{equation}
where $W_g$ represents the learnable parameters of the linear layer and $\text{Softmax}$ formula is given as follows:
\begin{equation}
  \text{Softmax}(z_i) = \dfrac{\exp(z_i)}{\sum_j{\exp(z_j)}}
\end{equation}

\textbf{Continuous Mixtures of Experts}. Mathematically, the continuous MoE model can be formulated as:
\begin{equation}
  \sum_{i=1}^n g(x)_i f_i(x) \label{eq:moe}
\end{equation}
\noindent where \( g(x)_i \) is the $i-$th component in the $n-$dimensional outputs, for the \( i \)-th expert \( f_i(x) \). Here, each expert $f_i$ receives the same input \( x \) but is weighted differently based on the gating output. The learnable gating function dynamically determines the contribution of each expert, assigning higher weights to more relevant experts. In this continuous MoE framework, all experts participate in the final aggregation, with their influence proportional to their assigned weights.

  {\bf Sparse Mixtures of Experts (SMoE)~\cite{shazeer2017outrageously}.} SMoE is a variant of MoE that selects only the top-K experts based on the gating function, reducing computational costs by routing the input through a subset of experts. The SMoE gating mechanism is defined as:
\begin{equation}
  G(x):=\text{Softmax}(\text{TopK}(x \cdot W_g)) \label{eq:smoe-gate}
\end{equation}
where $(\text{TopK}(l))_i:=l_i$ if $l_i$ is within the top-K values $l \in \mathbb{R}^n$ and $(\text{TopK}(l))_i:=-\infty$ otherwise. This ensures that only the most relevant experts receive nonzero gating weights after applying the softmax function.

The final output of SMoE is then computed as:
\begin{equation}
  \sum_{i=1}^n G(x)_i f_i(x) \label{eq:smoe}
\end{equation}
Top-K experts selection in SMoE maintains model expressiveness while significantly improving efficiency.




\subsection{Composition of Variant Experts via Hidden Factors (\proposedcove$_h$)}

Here, we introduce the Composition of Variant Experts (\proposedcove), which differs from the traditional Mixture-of-Experts approaches in several key ways.
First, instead of employing multiple instances of identical feed-forward network (FFN), \proposedcove\ leverages different types of experts. Particularly, these experts can include various recommendation models, such as collaborative filtering models (e.g., BPR~\cite{rendle2009bpr}) for capturing user preferences and sequential models (e.g., GRU4Rec~\cite{gru4rec1}) for modeling session-based interactions.
Second, each expert interprets the same input data differently, offering diverse perspectives. For example, BPR treats input data as user-item interactions, while GRU4Rec models sequential behavior.
Third, because each expert digests input data differently, during training, all experts are trained simultaneously to give all experts the ability to generalize; during inference, only the most relevant experts are activated.
Fourth, we standardize the Compositions of Variant Experts so that each expert can determine its own input to the routing gate, which is then aggregated to produce the input to the gating mechanism.
More details will be discussed in the next paragraphs.

\textbf{Standardized Variant Experts.} Various type of experts may produce various type of outputs from various type of inputs, providing unique insights into the data. For example, BPR learns from triplet user-positive-negative item interactions, optimizing rankings to favor positive items, providing robust user and item representations. In addition, \gruforrec\ captures temporal user behaviors by predicting the next item in a sequence of interactions, relying sorely on item sequence for input and optimizing the ranking of the next item accordingly. By integrating diverse expert models, we achieve a more comprehensive understanding of user behavior and preferences.



Despite their differences, we standardize the output of all experts including three components:

\begin{itemize}
  \item Hidden context representation: $d-$dimensional vector that captures the user $u$ context up to time $T$
        \begin{equation}
          h_i(u, T) = \Phi_i\left( u, C^u_{1:T} \right) \in \mathbb{R}^d
        \end{equation}
  \item Item embeddings: $\Psi(p) \in \mathbb{R}^d$ is a $d-$dimensional vector representation of item $p$
  \item Item bias: $\beta(p) \in \mathbb{R}$ is a scalar representing the bias of item $p$
\end{itemize}

Each expert serves as a sub-model. Using a gating mechanism, the three components from all experts are aggregated to form the final representations. We named this approach \proposedcove$_h$ as the aggregated components are hidden representation from variant of experts.


\textbf{Gating Mechanism.} The gating mechanism determines the relevance of each expert based on the input data, enabling the model to dynamically select the most appropriate experts for a given context. More concretely, the contribution of each expert is decided upon the model receiving the inputs via the gating mechanism. By learning the contribution of each expert, the model can adapt to diverse user behaviors and preferences, ultimately enhancing recommendation quality.

Each of the $n$ experts determines its own $d-$dimensional input to the gating mechanism, denoted as $e_i(u, C^u_{1:T})$. These individual inputs are then aggregated to form an $n \times d-$dimensional input for the gate. This design allows for flexibility in selecting the most relevant features for the gating mechanism, tailored to each expert. For instance, given a specific input, user representations might be crucial for the gate in BPR, whereas GRU4Rec may utilize embedded input items as gate inputs. Once computed, all experts' gate inputs are aggregated and passed through the gating mechanism.

In this work, we employ the simple gate mechanism consisting of a fully connected layer with a softmax activation function. This ensures that the gate assigns a weight to each expert while maintaining a total sum of $1$. Particularly, the gating mechanism can be represented as:
\begin{equation}
  g\left(u,C^u_{1:T}\right) := \text{Softmax}\left( W_g \left[e_i\left(u,C^u_{1:T}\right) \Big|_1^n \right] \right)
  \label{eq:gate}
\end{equation}

\noindent where \( e_i\left(u,C_{1:T}^u\right) \) is the \( i \)-th expert's gate input, \(\left[e_i\left(u,C_{1:T}^u\right) \Big|_1^n\right]\) denotes the concatenation of all gate inputs, and $W_g$ is the trainable weights.
The output of gate $g\left(u,C^u_{1:T}\right) \to \mathbb{R}^n$ is subsequently used to weigh the contributions of the experts' outputs to the final representation.

\textbf{Training Variant Experts.}
Our \proposedcove\ framework enables each expert to process input data differently. During training, all experts are optimized simultaneously to collectively provide a comprehensive view of the data. This approach ensures that each expert learns distinct aspects of the data, contributing to the final recommendation.



The gating mechanism determines the contribution of each expert based on the input data. The final hidden representation, item embedding, and bias term are computed as weighted sums of the corresponding components from all experts, where the weights are determined by the gating function:
\begin{align}
  h(u)     & = \sum_{i=1}^{n} g\left(u,C^u_{1:T}\right)_i \times h_i(u,T)   \\
  \Psi(p)  & = \sum_{i=1}^{n} g\left(u,C^u_{1:T}\right)_i \times \Psi_i(p)  \\
  \beta(p) & = \sum_{i=1}^{n} g\left(u,C^u_{1:T}\right)_i \times \beta_i(p)
\end{align}
where \(g(u,C^u_{1:T})\) is derived from ~\autoref{eq:gate}, and $h_i(u,T), \Psi_i(p), \beta_i(p)$ represent the hidden context, item embedding, and item bias deriving from the $i-$th expert.

The final item score for a user $u$ and item $p$ is computed by taking the dot product between the hidden representation and item embedding, and then adding the item bias:
\begin{equation}
  \hat{{r}}_{u,p} = h(u) \cdot \Psi(p) + \beta(p)
\end{equation}


\begin{figure*}[htb]
  \centering
  \includegraphics[width=0.625\textwidth]{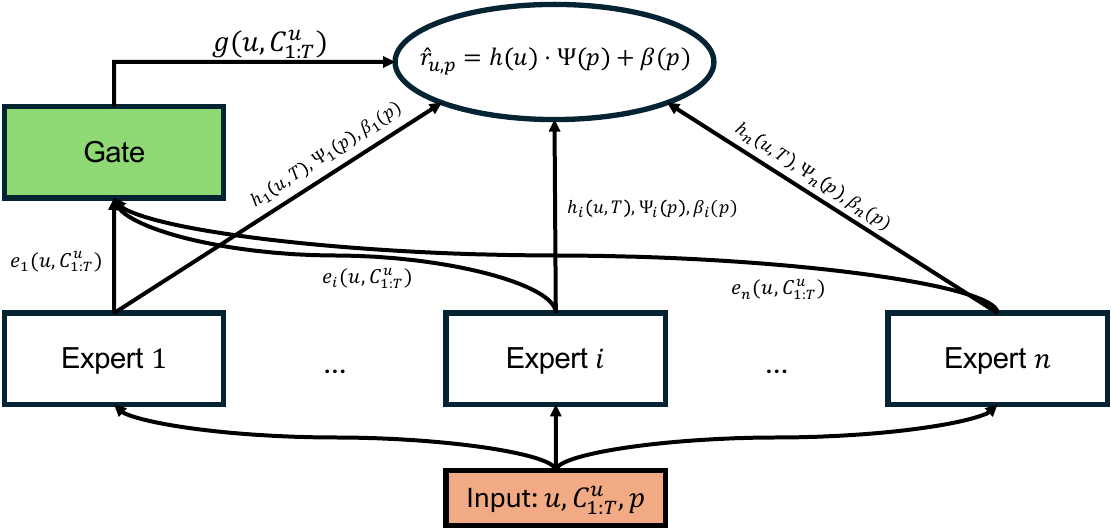}
  \caption{Overall flow of \proposedcove$_h$ variant. All experts receive user identity $u$ with their historical interactions $C_{1:T}^u$ and candidate item $p$. Each expert $i$ then returns the current context representation $h_i(u,T)$, item embeddings $\Psi_i(p)$, and item biases $\beta_i(p)$.}
  \label{fig:move}
\end{figure*}

{\bf Learning objective.} The general learning objective is to enhance ranking performance between positive and negative items in the candidate set. Its primary objective is to maximize the probability that positive items are ranked higher than negative ones. This can be achieved by minimizing the negative log-likelihood:
\begin{equation}
  \ell = - \frac{1}{N_I} \log \sigma(\hat{r}_i - \hat{r}_j)
\end{equation}
\noindent where $\hat{r}_i$ and $\hat{r}_j$ denote the predicted scores for a positive item (a preferred item) and a negative item (an unobserved or less preferred item), respectively, and $N_I$ is the total number of item pairs. This formulation treats all negative items equally, ensuring uniform optimization across the training samples.

In practice, however, the learning dynamics differ between popular and long-tail items. Popular items are frequently sampled and updated during training, while long-tail items (those with fewer interactions) receive less attention. As a result, the prediction scores for long-tail items tend to remain low, making them ``easy negatives'' that do not significantly impact the optimization process. By contrast, distinguishing between positive items and popular negative items is more challenging.

To address this imbalance, BPR-max loss incorporates the softmax score corresponding to each negative item in the candidate set so that it assigns lower weights to these easy negatives, allowing the model to focus more on harder negatives, minimizing the following loss:
\begin{equation}
  \ell = - \log \sum_i\sum_{j=1}^{N_I} s_j \sigma(\hat{r}_i - \hat{r}_j)
\end{equation}
where $s_j = \frac{\exp(\hat{r}_j)}{\sum_{k=1}^{N_I}{\exp(\hat{r}_k)}}$ is the normalization score of the negative item $j$ via the Softmax function. By prioritizing these challenging comparisons, the model promotes more effective learning and improves its ability to rank positive items accurately across a diverse set of candidates.


  {\bf Inference.} In our unique framework, each expert observes input data differently, so in the training phase, we employed the continuous gate mechanism. During inference, we derive the final prediction using sparse gate mechanism, where only the most relevant experts are activated based on the input data. This routing mechanism allows the model to dynamically select the most appropriate experts for the given context, reducing the computational cost since there only $k$ experts were selected. We entrust the routing mechanism to choose the best experts. Following Equation~\ref{eq:smoe}, the hidden representation during inference is computed as:
\begin{equation}
  h(u) = \sum_{i=1}^n G(u,C^u_{1:T})_i f_i(u,C^u_{1:T})
  \label{eq:inference}
\end{equation}
\noindent where $G(u,C^u_{1:T})$ is the sparse gate defined in~\autoref{eq:smoe-gate}.




\begin{figure*}[htb]
  \centering
  \includegraphics[width=0.625\textwidth]{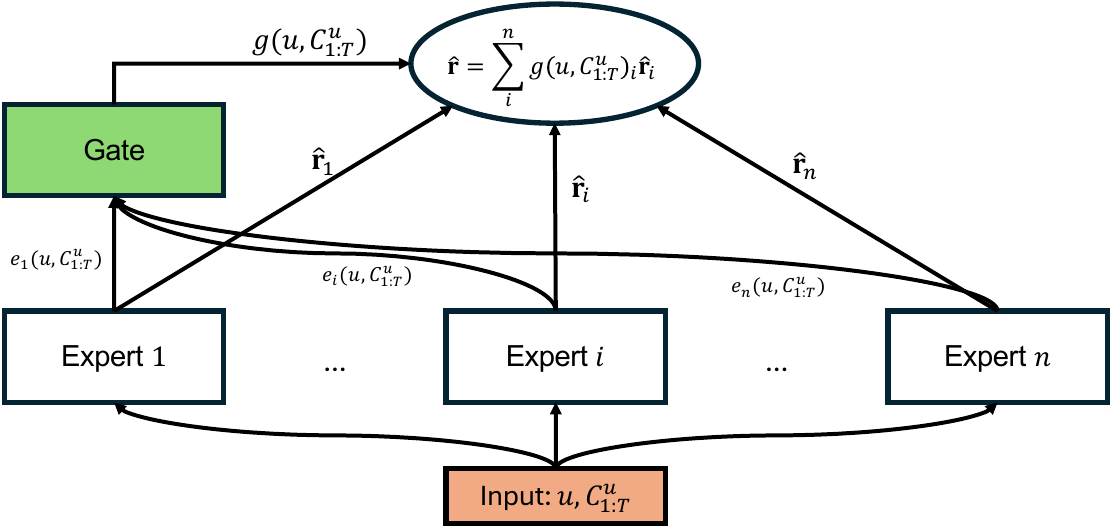}
  \caption{Overall flow of \proposedcove$_s$ variant. All experts receive user identity $u$ with their historical interactions $C_{1:T}^u$ and candidate item $p$. Here, each expert $i$ just returns its predicted scores for the set of candidate items $\hat{\mathbf{r}}_i$.}
  \Description{Overall flow of \proposedcove$_s$ variant. All experts receive user identity $u$ with their historical interactions $C_{1:T}^u$ and candidate item $p$. Here, each expert $i$ just returns its predicted scores for the set of candidate items $\hat{\mathbf{r}}_i$.}
  \label{fig:mose}
\end{figure*}

\subsection{Composition of Variant Experts via Scoring Function (\proposedcove$_s$)}
We propose another approach for aggregation that offers greater flexibility but shallower expert connections. Unlike \proposedcove$_h$ which experts are combined at \textit{hidden} representation level, this approach, referred to as \proposedcove$_s$, gathers item \textit{scores} predictions from individual experts.

Specifically, each expert generates predicted item scores directly, which are subsequently combined using a weighted sum via the same routing mechanism. This method offers greater flexibility compared to the \proposedcove$_h$ approach, as it does not require experts to share the same hidden dimensions, provided that an expert is capable of providing scores over the item set. This approach bears resemblance to a learnable ensemble method, yet it possesses the advantage of trainable aggregation weights that can adapt dynamically to varied contexts. Consequently, the application of the same set of experts may result in varying weights, depending on the specific context.

The final ranking scores for all items are formally computed using the gating aggregation as follows:

\begin{equation}
  \hat{\mathbf{r}} = \sum_{i=1}^{N} g\left(u,C^u_{1:T}\right)_i \hat{\mathbf{r}}_i
  \label{eq:mose}
\end{equation}

\noindent where \( \hat{\mathbf{r}}_i \in \mathbb{R}^N \) denotes the scores of all items provided by the \( i \)-th expert.



\section{Related Work}\label{sec:related}
In this section, we first survey related work that deals with general personalized recommendation (learning long-term preferences) and sequential recommendation (learning short-term preferences). In addition, we also survey existing LLM approaches for recommendations.
Lastly, we describe related mixture-of-experts.

{\bf Personalized Recommendation.}
The general recommender systems produce a list of suggested items for a target user. A crucial part is to model long-term user preferences for personalized recommendations. Many studies are based on Matrix Factorization~\cite{mnih2007pmf,hu2008collaborative,koren2009matrix,rendle2009bpr} that leverage past user-item interactions. In addition, many others try to address the sparsity issue using auxiliary information such as social networks~\cite{ma2008sorec}, topic modeling~\cite{wang2013collaborative},  reviews~\cite{zheng2017joint,tay2018multi}, questions~\cite{le2024question}, images~\cite{chen2019personalized}, graphs~\cite{he2015trirank,jendal2024hypergraphs}. Modern online~\cite{he2016fast} or continual learning~\cite{do2023continual} approaches also leverage collaborative filtering as the backbone models.
Within the scope of this work, we evaluate our framework with user-item interaction data only. The basis of the recommendation optimization objective is to estimate missing ratings by minimizing the reconstruction of observed scores, e.g., MF~\cite{koren2009matrix}, or infer ranking scores by maintaining the favoring of positive items over negative items, e.g., BPR~\cite{rendle2009bpr}. In this work, we concentrate on the latter, which is more suitable for implicit feedback data (e.g., clicks, views).

  {\bf Sequential Recommendation.}
Sequential recommendation (SR), which aims to predict a user's next action based on their historical temporal interactions, has attracted significant attention in recent years. This is particularly fueled by the rise of deep learning techniques and the growing relevance of sequential patterns in modern user behaviors.
Sequential recommendation has evolved into various taxonomies, including $(i)$ pure ID-based SR, $(ii)$ multi-modal SR, $(iii)$ cross-domain SR, and $(iv)$ LLM-powered SR. Of particular interest to this work is the pure ID-based SR, which focuses on modeling order and dependency of user-item interactions without additional information such as text, images, or other modalities.
Prior works have explored various approaches to model sequential data, ranging from traditional methods such as Markov Chains~\cite{rendle2010factorizing,wang2015learning} and translation-based~\cite{he2017translation}; modeling repeating behavior~\cite{bhagat2018buy,ren2019repeatnet}; to deep learning, including recurrent neural networks (RNNs)~\cite{gru4rec1,hidasi2018recurrent}, Transformer-based architectures~\cite{sun2019bert4rec,kang2018self,de2021transformers4rec}.
FPMC~\cite{rendle2010factorizing} subsumes matrix factorization with Markov Chains to model personalized next-basket recommendation. TransRec~\cite{he2017translation} assumes and learns a transition representation: item + user $\approx$ next item.
Leveraging the success of RNNs in sequential modeling, GRU4Rec~\cite{gru4rec1} uses gated recurrent unit (GRU) to model session data, capturing the temporal dynamics of user behavior. Based on the initial success, GRU4Rec+~\cite{hidasi2018recurrent} extends its predecessor by incorporating a novel negative sampling strategy and new ranking loss functions.
Convolutional neural networks (CNNs)~\cite{chen2022double,tang2018personalized,yan2019cosrec,yuan2019asimple} have also been applied to sequential recommendation to capture local patterns in user behavior, enabling faster parallel computation.
Graph-based approaches~\cite{wang2020next,wang2021session,yang2022multi} have been explored due to the flexibility in modeling complex social relationships and transition patterns using graph neural networks.
With the advent of attention and Transformer-based architectures~\cite{vaswani2017attention}, sequential recommendation has seen significant advancements~\cite{kang2018self,sun2019bert4rec,ying2018sequential,tanjim2020attentive,petrov2022recencysampling,wang2023sequential}.
NARM~\cite{li2017neural} and STAMP~\cite{liu2018stamp} utilize attention mechanisms to model the importance of different items \textit{within the same session}, creating \textit{long-term} or \textit{local} representation, which is then combined with the last hidden factors, referred to as \textit{short-term} or \textit{global} feature. These long-term and short-term features differ from our definitions of short- and long-term preference models in~\autoref{sec:short-vs-long-term-preferences}.
Transformer models, such as SASRec~\cite{kang2018self} and BERT4Rec~\cite{sun2019bert4rec}, have been widely adopted due to their ability to capture long-range dependencies and parallelize computations. Transformers4Rec~\cite{de2021transformers4rec} facilitates the use of Transformer-based architectures in sequential recommendation by providing a unified library for various Transformer-based models, using HuggingFace Frameworks~\cite{wolf-etal-2020-transformers}. To this date, Transformers-based models and their variants remain one of the most popular and effective approaches for sequential recommendation~\cite{petrov2025efficient,petrov2023gsasrec,petrov2025improving,abbattista2024enhancing}.

Despite the popularity, recent studies on the reproducibility of recommender systems have shown that many models do not achieve their full potential due to issues with $(i)$ implementation correctness~\cite{petrov2022systematic,hidasi2023effect,milogradskii2024revisiting}, $(ii)$ hyper-parameter tuning~\cite{shehzad2023everyone}, or $(iii)$ the use of different loss functions~\cite{hidasi2018recurrent,klenitskiy2023turning}.
\cite{petrov2022systematic} have done hypothesis testing and concluded that the ``superior'' BERT4Rec is not systematically better than SASRec in the published literature, and that the performance gap is mainly due to the poor replicability. \cite{klenitskiy2023turning} shows that the performance gap between SASRec and BERT4Rec is due to the use of different loss functions, where SASRec uses binary cross-entropy (BCE) loss and BERT4Rec is trained using cross-entropy (CE) loss. They further show that the performance of SASRec can be improved and outperform BERT4Rec by using CE loss.
\cite{de2021transformers4rec,petrov2022systematic,klenitskiy2023turning} also show that BPR~\cite{rendle2009bpr} and GRU4Rec~\cite{gru4rec1,hidasi2018recurrent} are simple yet strong baselines for sequential recommendation, and can achieve competitive results with BERT4Rec~\cite{sun2019bert4rec} with proper negative sampling and loss function.

Other taxonomies of sequential recommendation, including multi-modal and cross-domain, are not the focus of this study. We refer to~\cite{chen2024survey,li2024graph,nasir2023survey,chen2023survey} for more comprehensive surveys.


  {\bf LLM for Recommendations.} The utilization of a large language model (LLM) in recommendation systems has garnered significant attention in recent research. Language model architectures have been extensively adopted in recommendation systems due to their inherent similarities in sequential characteristics, particularly in sequential recommendation tasks, as evidenced by models such as SASRec~\cite{kang2018self} and BERT4Rec~\cite{sun2019bert4rec}. The incorporation of large language models (LLMs) into recommendation systems is a growing trend. This integration is primarily driven by the LLMs' capacity for textual understanding, which enables text-based recommendations and the incorporation of textual modalities into recommendation systems~\cite{geng2022recommendation,ren2024representation,kim2025lost}.
However, their role in ID-based recommendation remains limited, often functioning as data augmentation component rather than a core model~\cite{chen2025enhancing}.
There is no clear evidence showing that LLM can easily outperform traditional models in ID-based recommendation~\cite{hou2023learning,yuan2023go,li2023exploring}.
On the other hand, the high computational cost and input token limitations for datasets with large number of items outweigh their performance benefits, making their integration a challenging trade-off.
We respectfully clarify that most LLM-based recommendation methods and our work address fundamentally different problem settings. These methods typically focus on \emph{multimodal recommendation} where rich textual information (item descriptions, user reviews, product metadata) is available, leveraging LLMs' natural language understanding capabilities to process this textual data. In contrast, our work focuses on \emph{pure ID-based sequential recommendation} where only interaction sequences (user IDs, item IDs, timestamps) are available, without any textual content. This is common in scenarios where textual data is unavailable, privacy regulations prohibit using textual user data, or real-time systems require low-latency inference. These represent different research challenges with different applicable scenarios, making direct comparison unfair without either adding textual features to our datasets (changing our problem setting) or removing textual features from LLM-based methods (undermining their design). Among LLM-based approaches, only methods that operate purely on user and item IDs, such as OpenP5~\cite{xu2024openp5}, provide a fair basis for comparison in our pure ID-based setting.



  {\bf Mixture-of-Experts} models have recently garnered significant attention, particularly following the success of~\cite{jiang2024mixtral}, with various applications in natural language processing. The existing Mixture-of-Experts for recommendation research usually tackles multi-task learning formulation~\cite{ma2018modeling,xu2024MoME}, where the shared bottom networks contain general information while experts are trained for specific tasks. The routing mechanism will then select suitable experts for the task of given input.
Orthogonally,~\cite{bian2023multimodal} proposes a mixture of three experts for multi-modal sequential recommendation, including text, ID, and image experts. Comparisons with
these MoE models are beyond the scope of this study.

\section{Experiments}\label{sec:exp}
In this section, we evaluate our proposed \proposedcove\ in the task of session-aware recommendation, mainly on two publicly available benchmark datasets. The improvement in ranking performance validates the effectiveness of our solution.

\subsection{Experimental Setup} \label{subsec:experiment-setup}

\textbf{Datasets.}
Recent studies have identified several widespread flaws in recommendation systems, including dataset-task mismatch and negative sampling during testing, which hinder accurate evaluation~\cite{tang2018personalized,krichene2020sampled,hidasi2023widespread}. In this paper, we strive to follow best practices to enhance the effectiveness of model evaluation.

To address the dataset-task mismatch, we utilize three session-aware datasets\footnote{In contrast, previous studies converted user ratings into session data for evaluation. However, since these rating-based datasets were not inherently session-specific and may exhibit weaker sequential signals~\cite{tang2018personalized}, they were excluded from this study.}, Diginetica, RetailRocket, and Cosmetics. In these datasets, each user interacts with multiple items within a browsing \emph{session}, and may have several such sessions recorded.

For statistical sufficiency, we retain users with at least three sessions and items with at least five interactions. The last two sessions of each user are reserved for validation and testing, while the remaining sessions are used for training. These two reserved sessions are randomly assigned to validation and test sets. The dataset statistics are presented in Table~\ref{tab:data_stats}. Code and data for reproducibility are publicly available\footnote{\url{https://github.com/PreferredAI/CoVE}}.

Furthermore, during the evaluation phase, we score and rank items across the entire item set rather than relying on candidate sampling. This approach mitigates biases introduced by the selection of candidate sets~\cite{krichene2020sampled,canamares2020target,dallmann2021case}, ensuring fairer comparisons and improved reproducibility.

\begin{table*}[t]
  \centering
  \caption{Data statistics}\label{tab:data_stats}
  \begin{tabular}{lrrr}
    \toprule
    ~                          & Diginetica & RetailRocket & Cosmetics
    \\ \midrule
    \#interactions             & 12,146     & 230,817      & 2,533,262
    \\
    \#users                    & 571        & 4,249        & 17,268
    \\
    \#sessions                 & 2,670      & 24,732       & 172,242
    \\
    \#items                    & 6,008      & 36,658       & 42,367
    \\
    \#sessions per user        & 4.68       & 5.82         & 9.97
    \\
    \#interactions per item    & 2.02       & 6.30         & 59.79
    \\
    \#interactions per session & 4.55       & 9.33         & 14.71
    \\
    density                    & 0.354\%    & 0.148\%      & 0.346\%
    \\ \bottomrule
  \end{tabular}
\end{table*}

{\bf Evaluation metrics.} To evaluate the effectiveness of each model, we let each model produce a ranked list of recommended items for every user. We quantitatively evaluate ranking performance using various well-known ranking metrics, including Area Under the ROC Curve (AUC), Mean Reciprocal Rank (MRR), Normalized Discounted Cumulative Gain (NDCG), and Recall. For these metrics, higher values indicate better quality. Comparisons between methods are tested with one-tailed paired-sample Student's t-test at $0.05$ level (i.e., $p\text{-value}<0.05$).

\textbf{Expert Selection.}
As mentioned in~\autoref{sec:related}, sequential recommendation models may not achieve their full replicability and reproducibility due to flaws in implementation and hyper-parameter tuning~\cite{petrov2022systematic,hidasi2023effect,milogradskii2024revisiting,shehzad2023everyone}; while simple yet effective models such as BPR~\cite{rendle2009bpr} and GRU4Rec~\cite{gru4rec1,hidasi2018recurrent} can achieve competitive results with proper training settings~\cite{de2021transformers4rec,petrov2022systematic,klenitskiy2023turning}.
Furthermore, \cite{ludewig2019performance,latifi2021session} have demonstrated that shallow models can outperform deep models in various settings.
In our experiments, we not only focus on deep models but also on popular and effective shallow models as experts, including two short-term preference models and two long-term preference models:

\begin{itemize}
  \item Short-term preference models as expert:
        \begin{itemize}
          \item \textbf{GRU4Rec}~\cite{gru4rec1,hidasi2018recurrent}\footnote{In our experiment, we use the improved GRU4Rec model proposed in \cite{hidasi2018recurrent}.}: GRU4Rec is a session-based recommendation model that employs a Gated Recurrent Unit (GRU) to capture sequential user behavior. Benchmarks indicate that GRU4Rec outperforms more recent session-based recommendation models.
          \item \textbf{SASRec++}: SASRec~\cite{kang2018self} is a session-based recommendation model that leverages self-attention mechanisms to capture user preferences. The SASRec+~\cite{klenitskiy2023turning} variant using Cross-Entropy loss can outperform SASRec with Binary Cross-Entropy loss and the allegedly superior BERT4Rec~\cite{sun2019bert4rec}. In this paper, we employ the BPR-max loss for SASRec, named SASRec++, which yields the best performance.
        \end{itemize}

  \item Long-term preference models as expert:
        \begin{itemize}
          \item \textbf{BPR}~\cite{rendle2009bpr}: BPR is a collaborative filtering model that optimizes the ranking of positive items over negative ones.
                BPR is a simple yet effective model widely used in recommendation systems.
          \item \textbf{FPMC~\cite{rendle2010factorizing}}: Factorized Personalized Markov Chains is a sequential recommendation model that captures user preferences by modeling item transitions within a session.
        \end{itemize}
\end{itemize}

\textbf{Comparative Methods.} In this work, we focus exclusively on the pure ID-based sequential recommendation problem, where models rely solely on user and item identifiers without leveraging contextual data such as textual descriptions, visual content, or other side information. To comprehensively evaluate the effectiveness of our proposed method, we compare \proposedcove$_h$ and \proposedcove$_s$ models with individual selected short-term and long-term preference models as well as the following baselines:

\begin{itemize}
  \item \textbf{\pfive}~\cite{xu2024openp5} is a multi-task LLM-based recommendation model that leverages pure user and item IDs for various recommendation tasks. As mentioned in~\autoref{sec:related}, \pfive\ operates within the pure ID-based paradigm, making it a relevant baseline for comparison.
  \item \textbf{BERT4Rec}~\cite{sun2019bert4rec} is a session-based recommendation model that employs a Transformer architecture to capture sequential user behavior. BERT4Rec is included as a baseline due to its popularity and effectiveness in session-based recommendation.
  \item \textbf{LightGCN}~\cite{he2020lightgcn} is a graph-based recommendation model that captures user-item interactions through a lightweight Graph Convolutional Network (GCN). LightGCN is included as a baseline due to its effectiveness in capturing user-item relationships. LightGCN will also be used as a long-term preference model in our investigation of \proposedcove$_s$ with graph-based experts.
  \item \textbf{HGRU4Rec}~\cite{quadrana2017personalizing} is a hierarchical GRU-based method for session-aware recommendation which take long-term user preferences into account. HGRU4Rec is included as a baseline due to its similarity in modeling both short- and long-term preferences.
\end{itemize}

\textbf{Hyper-parameter Tuning.}
For all baselines, we perform a grid search with various hyper-parameters to find the best configuration. For \proposedcove$_h$ and \proposedcove$_s$, we also conduct a grid search to determine the optimal configuration. We use the validation set to tune the hyper-parameters to their best and the test set to evaluate model performance. The main hyper-parameters include:
\begin{itemize}
  \item \textbf{Loss Function.} We consider various loss functions, including BPR~\cite{rendle2009bpr}, BPR-max~\cite{hidasi2018recurrent}, binary cross-entropy loss~\cite{kang2018self}, cross-entropy~\cite{sun2019bert4rec}, top-1~\cite{gru4rec1}. Generally, BPR and BPR-max losses perform the best based on their ability to distinguish positive from negative samples.
  \item \textbf{Learning Rate.} We consider learning rates in the range of $[0.005, 0.1]$.
  \item \textbf{Batch Size.} We consider batch sizes in the range of $[32, 512]$. Different loss functions may require different batch sizes. For example, BPR-max loss fits well with a smaller batch size of around $[32,128]$, while BPR-loss can speed up with a larger batch size of 256 or 512 without trade-off in performance.
\end{itemize}

\textbf{Experts' Gate Input.} As described in Section~\ref{sec:method}, we give the freedom of deciding gate input to the experts. Individual experts can use their own unique features as gate input by a standardized function. In the experiments for \proposedcove$_h$ and \proposedcove$_s$, we follow common practice in NLP research~\cite{jiang2024mixtral}, where the input to the gate is item embeddings.

\subsection{Overall Performance}

\begin{table}[t]
  \centering
  \caption{Overall ranking performance: Comparison of \proposedcove\ against baselines across datasets}
  \label{tab:general_res}
  \begin{tabular}{llccccccccc}
    \toprule
     & \multirow{2}{*}{Model} & \multirow{2}{*}{AUC}        &  & \multirow{2}{*}{MRR} &  & \multicolumn{2}{c}{NDCG@k} &                      & \multicolumn{2}{c} {Recall@k}                                               \\\cmidrule{7-8}\cmidrule{10-11}
     &                        &                             &  &                      &  & k=10                       & k=20                 &                               & k=10                 & k=20
    \\
    \midrule
    \multirow{8}{*}{\rotatebox[origin=c]{90}{Diginetica}}
     & \gruforrec             & \underline{0.7771}          &  & \underline{0.2979}   &  & 0.3171                     & 0.3233               &                               & 0.3888               & 0.4133               \\
     & \sasrec                & 0.7666                      &  & 0.2943               &  & \underline{0.3301}         & \underline{0.3343}   &                               & \underline{0.4466}   & \underline{0.4641}   \\
     & BERT4Rec               & 0.7177                      &  & 0.3426               &  & 0.3560                     & 0.3575               &                               & 0.4011               & 0.4063               \\
     & BPR                    & 0.6334                      &  & 0.0748               &  & 0.0854                     & 0.0961               &                               & 0.1361               & 0.1775               \\
     & \fpmc                  & 0.6609                      &  & 0.0945               &  & 0.1107                     & 0.1205               &                               & 0.1805               & 0.2189               \\
     & LightGCN               & 0.7547                      &  & 0.2120               &  & 0.2488                     & 0.2601               &                               & 0.3835               & 0.4273               \\
     & \hgruforrec            & 0.5606                      &  & 0.1188               &  & 0.1333                     & 0.1392               &                               & 0.1893               & 0.2130               \\
     & \pfive                 & ---                         &  & 0.0027               &  & 0.0040                     & 0.0044               &                               & 0.0087               & 0.0105               \\
    \cmidrule{2-11}
     & \proposedcove$_s$      & 0.7854                      &  & \textbf{0.4112}$^\S$ &  & \textbf{0.4379}$^\S$       & \textbf{0.4416}$^\S$ &                               & \textbf{0.5254}$^\S$ & \textbf{0.5394}$^\S$ \\
     & \proposedcove$_h$      & \textbf{0.8007}$^\S$        &  & 0.3756$^\S$          &  & 0.4003$^\S$                & 0.4060$^\S$          &                               & 0.4869$^\S$          & 0.5096$^\S$          \\
    \cmidrule{2-11}
     & Improvement \%         & 3.04\%                      &  & 38.03\%              &  & 32.66\%                    & 32.10\%              &                               & 17.64\%              & 16.22\%
    \\
    \midrule
    \midrule
    \multirow{8}{*}{\rotatebox[origin=c]{90}{RetailRocket}}
     & \gruforrec             & \underline{0.8772}          &  & \underline{ 0.2687}  &  & \underline{0.3035}         & \underline{0.3175}   &                               & \underline{0.4356}   & \underline{0.4909}   \\
     & \sasrec                & 0.6606                      &  & 0.0962               &  & 0.1334                     & 0.1412               &                               & 0.2594               & 0.2892               \\
     & BERT4Rec               & 0.7827                      &  & 0.2278               &  & 0.2551                     & 0.2596               &                               & 0.3476               & 0.3653               \\
     & \bpr                   & 0.8035                      &  & 0.2161               &  & 0.2451                     & 0.2528               &                               & 0.3493               & 0.3794               \\
     & \fpmc                  & 0.8393                      &  & 0.2437               &  & 0.2744                     & 0.2848               &                               & 0.3879               & 0.4290               \\
     & LightGCN               & 0.8044                      &  & 0.1065               &  & 0.1238                     & 0.1333               &                               & 0.1970               & 0.2346               \\
     & \hgruforrec            & 0.8692                      &  & 0.2327               &  & 0.2612                     & 0.2731               &                               & 0.3735               & 0.4203               \\
     & \pfive                 & ---                         &  & 0.0018               &  & 0.0023                     & 0.0026               &                               & 0.0041               & 0.0055               \\
    \cmidrule{2-11}
     & \proposedcove$_s$      & 0.8767                      &  & 0.3365$^\S$          &  & 0.3675$^\S$                & 0.3783$^\S$          &                               & 0.4813$^\S$          & 0.5239$^\S$          \\
     & \proposedcove$_h$      & \textbf{0.8798}$^\S$        &  & \textbf{0.3406}$^\S$ &  & \textbf{0.3732}$^\S$       & \textbf{0.3831}$^\S$ &                               & \textbf{0.4928}$^\S$ & \textbf{0.5319}$^\S$ \\
    \cmidrule{2-11}
     & Improvement \%         & 0.30\%                      &  & 26.76\%              &  & 22.97\%                    & 20.66\%              &                               & 13.13\%              & 8.35\%
    \\
    \midrule
    \midrule
    \multirow{8}{*}{\rotatebox[origin=c]{90}{Cosmetics}}
     & \gruforrec             & 0.9147                      &  & \underline{0.1568}   &  & \underline{0.1842}         & \underline{0.2042}   &                               & \underline{0.3069}   & \underline{0.3857}   \\
     & \sasrec                & 0.9104                      &  & \underline{0.1568}   &  & 0.1823                     & 0.1999               &                               & 0.2967               & 0.3661               \\
     & BERT4Rec               & 0.8929                      &  & 0.1499               &  & 0.1737                     & 0.1906               &                               & 0.2803               & 0.3471               \\
     & \bpr                   & 0.8054                      &  & 0.0175               &  & 0.0189                     & 0.0250               &                               & 0.0396               & 0.0638               \\
     & \fpmc                  & \underline{\textbf{0.9248}} &  & 0.0985               &  & 0.1152                     & 0.1319               &                               & 0.2022               & 0.2684               \\
     & LightGCN               & 0.7342                      &  & 0.0037               &  & 0.0032                     & 0.0043               &                               & 0.0064               & 0.0111               \\
     & \hgruforrec            & 0.8948                      &  & 0.0764               &  & 0.0878                     & 0.1020               &                               & 0.1546               & 0.2113               \\
     & \pfive                 & ---                         &  & 0.0086               &  & 0.0194                     & 0.0200               &                               & 0.0111               & 0.0112               \\
    \cmidrule{2-11}
     & \proposedcove$_s$      & 0.9167$^\S$                 &  & 0.1643$^\S$          &  & 0.1937$^\S$                & 0.2135$^\S$          &                               & 0.3224$^\S$          & 0.4007$^\S$          \\
     & \proposedcove$_h$      & 0.9182$^\S$                 &  & \textbf{0.1647}$^\S$ &  & \textbf{0.1940}$^\S$       & \textbf{0.2139}$^\S$ &                               & \textbf{0.3229}$^\S$ & \textbf{0.4016}$^\S$ \\
    \cmidrule{2-11}
     & Improvement \%         & -0.71\%                     &  & 5.04\%               &  & 5.32\%                     & 4.75\%               &                               & 5.21\%               & 4.12\%
    \\
    \bottomrule
  \end{tabular}%
  \par\smallskip\small
  $^\S$: $p\text{-value}<0.05$ against the best-performing baseline. The best values are \textbf{bolded}. The best values among baselines are \underline{underlined}.
\end{table}

Here, we investigate whether using a mixture of experts enhances recommendation effectiveness compared to individual experts and baseline models across three datasets of varying scales: Diginetica, RetailRocket, and Cosmetics.

Table~\ref{tab:general_res} consistently shows that both variants of \proposedcove\ outperform the baselines across all three datasets in terms of MRR, NDCG, and Recall, with statistically significant improvements. The AUC metric, which measures the proportion of correctly ordered item pairs, does not fully capture recommendation quality in top-k scenarios. Nevertheless, \proposedcove$_h$ and \proposedcove$_s$ maintain competitive AUC results, close to or exceeding the best baseline performance.

  {\bf Performance across different dataset scales.} The results demonstrate that \proposedcove\ maintains its effectiveness across datasets of varying scales. On the smaller Diginetica dataset (12K interactions), \proposedcove$_s$ achieves the most substantial improvements, with 38.03\% improvement in MRR and 32.66\% improvement in NDCG@10 over the best baseline. On the medium-sized RetailRocket dataset (231K interactions), \proposedcove$_h$ shows the best overall performance with 26.76\% improvement in MRR. On the large-scale Cosmetics dataset (2.5M interactions, approximately 200 times larger than Diginetica), both variants continue to outperform baselines significantly, with \proposedcove$_h$ achieving 5.04\% improvement in MRR, 5.32\% in NDCG@10, and 5.21\% in Recall@10. The consistent improvements across different scales validate the robustness and scalability of our approach.

  {\bf Baseline analysis.} Among the individual expert models, \gruforrec\ remains the best-performing baseline overall, demonstrating its robustness in sequential recommendation tasks. The session-aware baseline \hgruforrec\ does not outperform \gruforrec, which aligns with findings in prior benchmarks~\cite{latifi2021session}.

Notably, the LLM-based \pfive\ model performs poorly across all datasets, suggesting that pure ID-based recommendation remains challenging for current LLM approaches.
As shown in the original research~\cite{xu2024openp5}, \pfive\ struggles and the performance degrades when the data becomes sparser (\#sessions/user, \#interactions/item, \#interactions/session). This observation is consistent with our findings, where \pfive\ fails to deliver competitive results on all three datasets. For instance, on Diginetica where the average number of sessions per user, interactions per item, and interactions per session are only 4.68, 2.02, and 4.55, respectively, \pfive\ achieves an MRR and NDCG@20 of merely 0.0027 and 0.0044.
The results improve slightly on the larger Cosmetics dataset, where these averages increase to 9.97, 59.79, and 14.71 respectively; however, \pfive\ still attains only an MRR of 0.0086 and NDCG@20 of 0.0200.
This highlights the limitations of LLMs in handling sparse recommendation data effectively.

It is worth noting that FPMC achieves the highest AUC on the Cosmetics dataset (0.9248), yet its performance on MRR, NDCG@k, and Recall@k metrics is significantly worse than \proposedcove. This observation can be explained by FPMC's pairwise ranking objective, which is designed to optimize AUC rather than top-k recommendation quality. FPMC itself can be interpreted as a combination of Matrix Factorization (MF) and Factorization Markov Chains (FMC), similar to having one long-term (MF) and one short-term (FMC) expert. However, unlike \proposedcove, each component in FPMC contributes equally toward the final optimization objective without adaptive weighting.

  {\bf Efficiency considerations.} A practical advantage of \proposedcove\ is the ability to leverage pretrained individual experts. Since each expert can be trained independently as a standalone model, we can use these pretrained experts to initialize \proposedcove, significantly reducing training time. This is particularly beneficial for large-scale datasets where training multiple experts altogether could be unstable. Additionally, during inference, we can employ a sparse gating mechanism using top-$k$ experts instead of continuous gating, which reduces computational costs while maintaining competitive recommendation quality.


\subsection{Ablation Study}\label{subsec:ablation-study}
In this section, we systematically study the contribution of different components to our \proposedcove. First, we investigate the effectiveness of the proposed gating mechanism. Then, we analyze various compositions of experts to gain insight in selecting the optimal combination of variant experts.

\begin{table}[t]
  \centering
  \caption{Gating effectiveness evaluation: Ranking performance of both variants of \proposedcove\ with and without gating mechanism}
  \label{tab:uniform_gate}
  \setlength{\tabcolsep}{4pt}
  \begin{tabular}{cccccccccccc}
    \toprule
    \multirow{2}{*}{Dataset}                             & \multirow{2}{*}{Model}             & \multirow{2}{*}{Gate} & \multirow{2}{*}{AUC} &                      & \multirow{2}{*}{MRR} &                      & \multicolumn{2}{c}{NDCG@k} &                      & \multicolumn{2}{c}{Recall@k}                                               \\\cmidrule{8-9}\cmidrule{11-12}
                                                         &                                    &                       &                      &                      &                      &                      & k=10                       & k=20                 &                              & k=10                 & k=20
    \\
    \midrule
    \multirow{4}{*}{\rotatebox[origin=c]{0}{Diginetica}} & \multirow{2}{*}{\proposedcove$_s$} & \checkmark            & \textbf{0.7854}      &                      & \textbf{0.4112}$^\S$ &                      & \textbf{0.4379}$^\S$       & \textbf{0.4416}$^\S$ &                              & \textbf{0.5254}$^\S$ & \textbf{0.5394}$^\S$ \\
                                                         &                                    & --                    & 0.7694               &                      & 0.3860               &                      & 0.4071                     & 0.4119               &                              & 0.4816               & 0.5009               \\
    \cmidrule{2-12}
                                                         & \multirow{2}{*}{\proposedcove$_h$} & \checkmark            & 0.8007               &                      & \textbf{0.3756}      &                      & \textbf{0.4003}            & \textbf{0.4060}      &                              & \textbf{0.4869}$^\S$ & \textbf{0.5096}      \\
                                                         &                                    & --                    & \textbf{0.8022}      &                      & 0.3694               &                      & 0.3939                     & 0.3997               &                              & 0.4799               & 0.5026
    \\
    \midrule\midrule
    \multirow{4}{*}{RetailRocket}                        &
    \multirow{2}{*}{\proposedcove$_s$}                   & \checkmark                         & \textbf{0.8767}$^\S$  &                      & \textbf{0.3365}$^\S$ &                      & \textbf{0.3675}$^\S$ & \textbf{0.3783}$^\S$       &                      & \textbf{0.4813}              & 0.5239                                      \\
                                                         &                                    & --                    & 0.8726               &                      & 0.2847               &                      & 0.3279                     & 0.3395               &                              & 0.4794               & \textbf{0.5251}      \\
    \cmidrule{2-12}
                                                         & \multirow{2}{*}{\proposedcove$_h$} & \checkmark            & 0.8826               &                      & \textbf{0.3419}$^\S$ &                      & \textbf{0.3735}$^\S$       & \textbf{0.3851}$^\S$ &                              & \textbf{0.4902}$^\S$ & \textbf{0.5357}      \\
                                                         &                                    & --                    & \textbf{0.8836}      &                      & 0.3216               &                      & 0.3549                     & 0.3670               &                              & 0.4782               & 0.5265               \\
    \midrule\midrule
    \multirow{4}{*}{Cosmetics}                           &
    \multirow{2}{*}{\proposedcove$_s$}                   & \checkmark                         & \textbf{0.9167}$^\S$  &                      & \textbf{0.1643}$^\S$ &                      & \textbf{0.1937}$^\S$ & \textbf{0.2135}$^\S$       &                      & \textbf{0.3224}$^\S$         & \textbf{0.4007}$^\S$                        \\
                                                         &                                    & --                    & 0.9104               &                      & 0.1498               &                      & 0.1745                     & 0.1932               &                              & 0.2871               & 0.3610               \\
    \cmidrule{2-12}
                                                         & \multirow{2}{*}{\proposedcove$_h$} & \checkmark            & \textbf{0.9182}$^\S$ &                      & \textbf{0.1647}$^\S$ &                      & \textbf{0.1940}$^\S$       & \textbf{0.2139}$^\S$ &                              & \textbf{0.3229}$^\S$ & \textbf{0.4016}$^\S$ \\
                                                         &                                    & --                    & 0.8465               &                      & 0.0823               &                      & 0.0948                     & 0.1076               &                              & 0.1602               & 0.2106               \\

    \bottomrule
  \end{tabular}%
  \par\smallskip\small
  $^\S$: $p\text{-value} < 0.05$. The best values are \textbf{bolded}.
\end{table}

\subsubsection{Gating Effectiveness}
To emphasize the contribution of the proposed gating mechanism, we evaluate the performance of both variants of \proposedcove\ with and without learnable weights. When the gate values are uniform (i.e., $g(x)_i=\frac{1}{K}$, where $K$ is the number of experts), all experts contribute equally toward the overall optimization objective. Table~\ref{tab:uniform_gate} reports the ranking performance comparison across all three datasets.

The results demonstrate that the learnable gating mechanism substantially improves recommendation quality. For \proposedcove$_s$, removing the learnable gate leads to consistent performance degradation across all datasets and metrics. On Diginetica, MRR drops from 0.4112 to 0.3860 (6.1\% decrease), while on RetailRocket and Cosmetics, the decreases are 15.4\% and 8.8\%, respectively. Similar patterns are observed for NDCG@k and Recall@k metrics, with statistically significant differences ($p<0.05$) in most cases.

For \proposedcove$_h$, the impact is even more pronounced, particularly on the large-scale Cosmetics dataset where uniform gating causes dramatic performance drops: MRR decreases by 50.0\% (from 0.1647 to 0.0823), NDCG@10 by 51.1\%, and Recall@10 by 50.4\%. On smaller datasets, the degradation is more moderate but still substantial, with MRR reductions of 1.6\% on Diginetica and 5.9\% on RetailRocket. Interestingly, uniform gating slightly improves AUC in some cases, but these differences are not statistically significant and do not translate to better top-k recommendation quality.

These results confirm that adaptively weighting expert contributions based on input context is crucial for \proposedcove's effectiveness. The uniform weighting strategy, which treats all experts equally regardless of context, fails to leverage the complementary strengths of different experts. This validates that balancing the contribution among variant experts is not trivial and requires learned dynamic gating.

\subsubsection{Same-Type Experts vs. Varying-Type Experts}
The results from~\autoref{sec:short-vs-long-term-preferences} indicate the presence of both short-term and long-term preferences within the session-aware recommendation context. We categorize expert compositions into two sub-categories: same-type experts (only short-term or only long-term) and varying-type experts (consisting both types). We systematically explore different compositions with 2, 3, and 4 experts. Results for the Diginetica dataset are presented in~\autoref{tab:diginetica-varying-experts}, while RetailRocket results are shown in~\autoref{tab:retail-varying-experts}.

{\bf Impact on \proposedcove$_h$.} On Diginetica, same-type expert combinations lead to severe performance degradation. Combining two short-term experts (GRU+SAS) yields MRR of only 0.1569, while two long-term experts (BPR+FPMC) achieve 0.2368, both substantially worse than the best individual expert GRU4Rec (0.2979). In contrast, varying-type combinations show dramatic improvements. The best 2-expert combination (GRU+FPMC) achieves MRR of 0.3756, a 26.1\% improvement over GRU4Rec alone. Similar patterns emerge on RetailRocket, where the varying-type combination (GRU+BPR) achieves MRR of 0.3406, a 26.8\% improvement over GRU4Rec's 0.2687, while same-type combinations underperform.

Interestingly, adding more experts does not always improve performance for \proposedcove$_h$. On Diginetica, the optimal configuration uses only 2 experts (GRU+FPMC), while 3-expert and 4-expert combinations show degraded performance. This suggests that \proposedcove$_h$ benefits most from a focused combination of one strong short-term and one strong long-term expert, where the gating mechanism can effectively balance their complementary strengths.

  {\bf Impact on \proposedcove$_s$.} For \proposedcove$_s$, the patterns differ significantly. On Diginetica, even same-type combinations achieve competitive results, where the two short-term experts (GRU+SAS) reach MRR of 0.4036, substantially outperforming the best individual expert. Varying-type combinations further improve performance, with the 3-expert configuration (GRU+SAS+BPR) achieving the best NDCG@10 (0.4385) and Recall@10 (0.5324). The 4-expert configuration achieves the highest MRR (0.4112), demonstrating that \proposedcove$_s$ can effectively leverage multiple experts.

On RetailRocket, \proposedcove$_s$ exhibits similar flexibility. The best 2-expert varying-type combination (GRU+BPR) achieves MRR of 0.3354, representing a 24.8\% improvement over GRU4Rec alone. \proposedcove$_s$ maintains strong performance even with 4 experts, suggesting its robustness in aggregating diverse expert predictions.

These results reveal fundamental differences between the two variants. \proposedcove$_h$, which combines experts at the hidden representation level, requires careful selection of complementary expert types and benefits from simpler 2-expert configurations. In contrast, \proposedcove$_s$, which aggregates expert scores, demonstrates greater flexibility and can effectively leverage multiple experts simultaneously. Both variants, however, consistently benefit from combining short-term and long-term experts rather than same-type combinations, validating our hypothesis that capturing both preference types is essential for effective session-aware recommendation.

\begin{table}[]
  \centering
  \caption{Ranking performance with varying different experts on Diginetica dataset. GRU and SAS are GRU4Rec and SASRec++ respectively.}
  \label{tab:diginetica-varying-experts}
  \setlength{\tabcolsep}{3pt}
  \begin{tabular}{ccccccccccccccccccc}
    \toprule
                                                                  & \multirow{4}{*}{\begin{tabular}{c}Number \\ of \\ Experts \\\end{tabular}} & \multicolumn{5}{c}{Expert model} &            & \multicolumn{9}{c}{Ranking metrics}
    \\ \cmidrule{3-7} \cmidrule{9-17}
                                                                  &                                                                            & \multicolumn{2}{c}{Short-term}   &            & \multicolumn{2}{c}{Long-term}       &            & \multirow{2}{*}{AUC} &  & \multirow{2}{*}{MRR} &  & \multicolumn{2}{c}{NDCG@k} &  & \multicolumn{2}{c}{Recall@k}                                                          \\ \cmidrule{3-4}\cmidrule{6-7}\cmidrule{13-14} \cmidrule{16-17}
                                                                  &                                                                            & GRU                              & SAS        &                                     & BPR        & FPMC                 &  &                      &  &                            &  & k=10                         & k=20            &  & k=10            & k=20            \\
    \midrule
    \multirow{13}{*}{\rotatebox[origin=c]{90}{\proposedcove$_h$}} & \multirow{6}{*}{2}                                                         & \checkmark                       & \checkmark &                                     &            &                      &  & 0.7287               &  & 0.1569                     &  & 0.1785                       & 0.1876          &  & 0.2627          & 0.2977          \\ 
                                                                  &                                                                            &                                  &            &                                     & \checkmark & \checkmark           &  & 0.7350               &  & 0.2368                     &  & 0.2730                       & 0.2781          &  & 0.3958          & 0.4168          \\ 
    \cmidrule{3-17}
                                                                  &                                                                            & \checkmark                       &            &                                     & \checkmark &                      &  & 0.7940               &  & 0.3507                     &  & 0.3720                       & 0.3803          &  & 0.4518          & 0.4851          \\
                                                                  &                                                                            &                                  & \checkmark &                                     & \checkmark &                      &  & 0.7921               &  & 0.2380                     &  & 0.2727                       & 0.2811          &  & 0.3940          & 0.4273          \\
                                                                  &                                                                            & \checkmark                       &            &                                     &            & \checkmark           &  & \textbf{0.8007}      &  & \textbf{0.3756}            &  & \textbf{0.4003}              & \textbf{0.4060} &  & \textbf{0.4869} & \textbf{0.5096} \\
                                                                  &                                                                            &                                  & \checkmark &                                     &            & \checkmark           &  & 0.6928               &  & 0.1321                     &  & 0.1504                       & 0.1562          &  & 0.2224          & 0.2452          \\
    \cmidrule{2-17}
                                                                  & \multirow{4}{*}{3}                                                         & \checkmark                       & \checkmark &                                     &            & \checkmark           &  & 0.7419               &  & 0.2189                     &  & 0.2341                       & 0.2444          &  & 0.2977          & 0.3380          \\

                                                                  &                                                                            & \checkmark                       & \checkmark &                                     & \checkmark &                      &  & 0.7800               &  & 0.2871                     &  & 0.3138                       & 0.3191          &  & 0.4063          & 0.4273          \\

                                                                  &                                                                            &                                  & \checkmark &                                     & \checkmark & \checkmark           &  & 0.7769               &  & 0.2377                     &  & 0.2708                       & 0.2785          &  & 0.3870          & 0.4168          \\

                                                                  &                                                                            & \checkmark                       &            &                                     & \checkmark & \checkmark           &  & 0.7806               &  & 0.3355                     &  & 0.3620                       & 0.3675          &  & 0.4536          & 0.4746          \\
    \cmidrule{2-17}
                                                                  & \multirow{1}{*}{4}                                                         & \checkmark                       & \checkmark &                                     & \checkmark & \checkmark           &  & 0.7744               &  & 0.3291                     &  & 0.3541                       & 0.3591          &  & 0.4413          & 0.4606          \\
    \midrule
    \midrule
    \multirow{13}{*}{\rotatebox[origin=c]{90}{\proposedcove$_s$}} & \multirow{6}{*}{2}                                                         & \checkmark                       & \checkmark &                                     &            &                      &  & 0.7930               &  & 0.4036                     &  & 0.4318                       & 0.4359          &  & 0.5236          & 0.5394          \\
                                                                  &                                                                            &                                  &            &                                     & \checkmark & \checkmark           &  & 0.7423               &  & 0.2502                     &  & 0.2844                       & 0.2893          &  & 0.4046          & 0.4238          \\ \cmidrule{3-17}
                                                                  &                                                                            & \checkmark                       &            &                                     & \checkmark &                      &  & \textbf{0.8119}      &  & 0.3674                     &  & 0.3845                       & 0.3968          &  & 0.4553          & 0.5044          \\
                                                                  &                                                                            &                                  & \checkmark &                                     & \checkmark &                      &  & 0.8059               &  & 0.3652                     &  & 0.3998                       & 0.4082          &  & 0.5166          & 0.5499          \\
                                                                  &                                                                            & \checkmark                       &            &                                     &            & \checkmark           &  & 0.8006               &  & 0.3521                     &  & 0.3726                       & 0.3801          &  & 0.4501          & 0.4799          \\
                                                                  &                                                                            &                                  & \checkmark &                                     &            & \checkmark           &  & 0.7943               &  & 0.3720                     &  & 0.4045                       & 0.4145          &  & 0.5166          & \textbf{0.5569} \\\cmidrule{2-17}
                                                                  & \multirow{4}{*}{3}                                                         & \checkmark                       & \checkmark &                                     &            & \checkmark           &  & 0.7860               &  & \textbf{0.4098}            &  & 0.4362                       & 0.4401          &  & 0.5219          & 0.5377          \\
                                                                  &                                                                            & \checkmark                       & \checkmark &                                     & \checkmark &                      &  & 0.7833               &  & 0.4097                     &  & \textbf{0.4385}              & \textbf{0.4428} &  & \textbf{0.5324} & {0.5499}        \\
                                                                  &                                                                            &                                  & \checkmark &                                     & \checkmark & \checkmark           &  & 0.7919               &  & 0.3578                     &  & 0.3980                       & 0.4023          &  & 0.5289          & 0.5464          \\
                                                                  &                                                                            & \checkmark                       &            &                                     & \checkmark & \checkmark           &  & 0.8035               &  & 0.3720                     &  & 0.3897                       & 0.4000          &  & 0.4606          & 0.5009          \\
    \cmidrule{2-17}
                                                                  & \multirow{1}{*}{4}                                                         & \checkmark                       & \checkmark &                                     & \checkmark & \checkmark           &  & 0.7854               &  & 0.4112                     &  & 0.4379                       & 0.4416          &  & 0.5254          & 0.5394          \\
    \bottomrule
    \hline
  \end{tabular}%
  \par\smallskip\small
  The best values are \textbf{bolded}.
\end{table}

\begin{table}[]
  \centering
  \caption{Ranking performance with varying experts on RetailRocket dataset.}
  \label{tab:retail-varying-experts}
  \setlength{\tabcolsep}{3pt}
  \begin{tabular}{ccccccccccccccccccc}
    \toprule
                                                                  & \multirow{4}{*}{\begin{tabular}{c}Number \\ of \\ Experts \\\end{tabular}} & \multicolumn{5}{c}{Expert model} &            & \multicolumn{9}{c}{Ranking metrics}
    \\ \cmidrule{3-7} \cmidrule{9-17}
                                                                  &                                                                            & \multicolumn{2}{c}{Short-term}   &            & \multicolumn{2}{c}{Long-term}       &            & \multirow{2}{*}{AUC} &  & \multirow{2}{*}{MRR} &  & \multicolumn{2}{c}{NDCG@k} &  & \multicolumn{2}{c}{Recall@k}                                                          \\ \cmidrule{3-4}\cmidrule{6-7}\cmidrule{13-14} \cmidrule{16-17}
                                                                  &                                                                            & GRU                              & SAS        &                                     & BPR        & FPMC                 &  &                      &  &                            &  & k=10                         & k=20            &  & k=10            & k=20            \\
    \midrule

    \multirow{13}{*}{\rotatebox[origin=c]{90}{\proposedcove$_h$}} & \multirow{6}{*}{2}                                                         & \checkmark                       & \checkmark &                                     &            &                      &  & 0.8782               &  & 0.3397                     &  & 0.3712                       & 0.3815          &  & 0.4869          & 0.5277          \\ 
                                                                  &                                                                            &                                  &            &                                     & \checkmark & \checkmark           &  & 0.7782               &  & 0.2221                     &  & 0.2515                       & 0.2595          &  & 0.3573          & 0.3886          \\ \cmidrule{3-17}
                                                                  &                                                                            & \checkmark                       &            &                                     & \checkmark &                      &  & 0.8798               &  & \textbf{0.3406}            &  & \textbf{0.3732}              & \textbf{0.3831} &  & \textbf{0.4928} & \textbf{0.5319} \\
                                                                  &                                                                            &                                  & \checkmark &                                     & \checkmark &                      &  & 0.8373               &  & 0.2224                     &  & 0.2484                       & 0.2593          &  & 0.3474          & 0.3904          \\
                                                                  &                                                                            & \checkmark                       &            &                                     &            & \checkmark           &  & \textbf{0.8853}      &  & 0.3393                     &  & 0.3716                       & 0.3823          &  & 0.4898          & 0.5321          \\
                                                                  &                                                                            &                                  & \checkmark &                                     &            & \checkmark           &  & 0.8212               &  & 0.1791                     &  & 0.1985                       & 0.2082          &  & 0.2777          & 0.3163          \\ \cmidrule{2-17}
                                                                  & \multirow{4}{*}{3}                                                         & \checkmark                       & \checkmark &                                     &            & \checkmark           &  & 0.8467               &  & 0.2942                     &  & 0.3218                       & 0.3316          &  & 0.4255          & 0.4639          \\
                                                                  &                                                                            & \checkmark                       & \checkmark &                                     & \checkmark &                      &  & 0.8822               &  & 0.3337                     &  & 0.3662                       & 0.3769          &  & 0.4853          & 0.5277          \\
                                                                  &                                                                            &                                  & \checkmark &                                     & \checkmark & \checkmark           &  & 0.7842               &  & 0.1600                     &  & 0.1800                       & 0.1880          &  & 0.2570          & 0.2885          \\
                                                                  &                                                                            & \checkmark                       &            &                                     & \checkmark & \checkmark           &  & 0.8680               &  & 0.3207                     &  & 0.3537                       & 0.3644          &  & 0.4740          & 0.5159          \\ \cmidrule{2-17}
                                                                  & \multirow{1}{*}{4}                                                         & \checkmark                       & \checkmark &                                     & \checkmark & \checkmark           &  & 0.8644               &  & 0.3126                     &  & 0.3439                       & 0.3545          &  & 0.4582          & 0.5001          \\
    \midrule\midrule
    \multirow{13}{*}{\rotatebox[origin=c]{90}{\proposedcove$_s$}} & \multirow{6}{*}{2}                                                         & \checkmark                       & \checkmark &                                     &            &                      &  & 0.8767               &  & 0.3365                     &  & 0.3675                       & 0.3783          &  & 0.4813          & 0.5239          \\
                                                                  &                                                                            &                                  &            &                                     & \checkmark & \checkmark           &  & 0.8150               &  & 0.2023                     &  & 0.2338                       & 0.2440          &  & 0.3490          & 0.3888          \\ \cmidrule{3-17}
                                                                  &                                                                            & \checkmark                       &            &                                     & \checkmark &                      &  & \textbf{0.8898}      &  & \textbf{0.3354}            &  & \textbf{0.3694}              & \textbf{0.3814} &  & \textbf{0.4952} & \textbf{0.5425} \\
                                                                  &                                                                            &                                  & \checkmark &                                     & \checkmark &                      &  & 0.8283               &  & 0.1820                     &  & 0.2404                       & 0.2524          &  & 0.4354          & 0.4825          \\
                                                                  &                                                                            & \checkmark                       &            &                                     &            & \checkmark           &  & 0.8877               &  & 0.3087                     &  & 0.3451                       & 0.3577          &  & 0.4792          & 0.5288          \\
                                                                  &                                                                            &                                  & \checkmark &                                     &            & \checkmark           &  & 0.8269               &  & 0.1489                     &  & 0.2036                       & 0.2179          &  & 0.3921          & 0.4481          \\\cmidrule{2-17}
                                                                  & \multirow{4}{*}{3}                                                         & \checkmark                       & \checkmark &                                     &            & \checkmark           &  & 0.8038               &  & 0.2593                     &  & 0.2824                       & 0.2949          &  & 0.3742          & 0.4239          \\
                                                                  &                                                                            & \checkmark                       & \checkmark &                                     & \checkmark &                      &  & 0.8235               &  & 0.2731                     &  & 0.3070                       & 0.3185          &  & 0.4297          & 0.4749          \\
                                                                  &                                                                            &                                  & \checkmark &                                     & \checkmark & \checkmark           &  & 0.8129               &  & 0.1560                     &  & 0.2129                       & 0.2261          &  & 0.4048          & 0.4566          \\
                                                                  &                                                                            & \checkmark                       &            &                                     & \checkmark & \checkmark           &  & 0.8828               &  & 0.3199                     &  & 0.3546                       & 0.3663          &  & 0.4815          & 0.5274          \\\cmidrule{2-17}
                                                                  & \multirow{1}{*}{4}                                                         & \checkmark                       & \checkmark &                                     & \checkmark & \checkmark           &  & 0.8840               &  & 0.2379                     &  & 0.2931                       & 0.3067          &  & 0.4841          & 0.5382          \\
    \bottomrule
  \end{tabular}%
  \par\smallskip\small
  The best values are \textbf{bolded}.
\end{table}

\subsection{\proposedcove$_h$ With Graph-Based Expert}

\begin{table}[]
  \centering
  \caption{\proposedcove$_h$ experimental results with graph-based expert. Best values are \textbf{bolded} and second best are \underline{underlined}.}
  \label{tab:w_graph}
  \setlength{\tabcolsep}{3pt}

  \begin{tabular}{ccccccccccccccccccc}
    \toprule
                                                            & \multirow{4}{*}{\begin{tabular}{c}Number \\ of \\ Experts \\\end{tabular}} & \multicolumn{5}{c}{Expert model} &            & \multicolumn{9}{c}{Ranking metrics}
    \\ \cmidrule{3-7} \cmidrule{9-17}
                                                            &                                                                            & \multicolumn{2}{c}{Short-term}   &            & \multicolumn{2}{c}{Long-term}       &            & \multirow{2}{*}{AUC} &  & \multirow{2}{*}{MRR} &  & \multicolumn{2}{c}{NDCG@k} &  & \multicolumn{2}{c}{Recall@k}                                                                   \\ \cmidrule{3-4}\cmidrule{6-7}\cmidrule{13-14} \cmidrule{16-17}
                                                            &                                                                            & GRU                              & SAS        &                                     & BPR        & LightGCN             &  &                      &  &                            &  & k=10                         & k=20               &  & k=10               & k=20               \\
    \midrule
    \multirow{8}{*}{\rotatebox[origin=c]{90}{Diginetica}}   & \multirow{4}{*}{1}                                                         & \checkmark                       &            &                                     &            &                      &  & {0.7771}             &  & {0.2979}                   &  & 0.3171                       & 0.3233             &  & 0.3888             & 0.4133             \\
                                                            &                                                                            &                                  & \checkmark &                                     &            &                      &  & 0.7666               &  & 0.2943                     &  & {0.3301}                     & {0.3343}           &  & {0.4466}           & {0.4641}           \\
                                                            &                                                                            &                                  &            &                                     & \checkmark &                      &  & 0.6334               &  & 0.0748                     &  & 0.0854                       & 0.0961             &  & 0.1361             & 0.1775             \\
                                                            &                                                                            &                                  &            &                                     &            & \checkmark           &  & 0.7547               &  & 0.2120                     &  & 0.2488                       & 0.2601             &  & 0.3835             & 0.4273             \\
    \cmidrule{2-17}
                                                            & \multirow{4}{*}{2}                                                         & \checkmark                       &            &                                     & \checkmark &                      &  & \underline{0.7940}   &  & \textbf{0.3507}            &  & \textbf{0.3720}              & \textbf{0.3803}    &  & \underline{0.4518} & \underline{0.4851} \\ 
                                                            &                                                                            &                                  & \checkmark &                                     & \checkmark &                      &  & {0.7921}             &  & 0.2380                     &  & 0.2727                       & 0.2811             &  & 0.3940             & 0.4273             \\ 
    \cmidrule{3-17}
                                                            &                                                                            & \checkmark                       &            &                                     &            & \checkmark           &  & \textbf{0.8139}      &  & 0.2870                     &  & 0.3156                       & 0.3245             &  & 0.4203             & 0.4553             \\ 
                                                            &                                                                            &                                  & \checkmark &                                     &            & \checkmark           &  & 0.7829               &  & \underline{0.3036}         &  & \underline{0.3429}           & \underline{0.3489} &  & \textbf{0.4711}    & \textbf{0.4956}    \\ 
    \midrule\midrule
    \multirow{8}{*}{\rotatebox[origin=c]{90}{RetailRocket}} & \multirow{4}{*}{1}                                                         & \checkmark                       &            &                                     &            &                      &  & \underline{0.8772}   &  & { 0.2687}                  &  & {0.3035}                     & {0.3175}           &  & {0.4356}           & \underline{0.4909} \\
                                                            &                                                                            &                                  & \checkmark &                                     &            &                      &  & 0.6606               &  & 0.0962                     &  & 0.1334                       & 0.1412             &  & 0.2594             & 0.2892             \\
                                                            &                                                                            &                                  &            &                                     & \checkmark &                      &  & 0.8035               &  & 0.2161                     &  & 0.2451                       & 0.2528             &  & 0.3493             & 0.3794             \\
                                                            &                                                                            &                                  &            &                                     &            & \checkmark           &  & 0.8044               &  & 0.1065                     &  & 0.1238                       & 0.1333             &  & 0.1970             & 0.2346             \\
    \cmidrule{2-17}
                                                            & \multirow{4}{*}{2}                                                         & \checkmark                       &            &                                     & \checkmark &                      &  & \textbf{0.8798}      &  & \textbf{0.3406}            &  & \textbf{0.3732}              & \textbf{0.3831}    &  & \textbf{0.4928}    & \textbf{0.5319}    \\ 
                                                            &                                                                            &                                  & \checkmark &                                     & \checkmark &                      &  & 0.8373               &  & 0.2224                     &  & 0.2484                       & 0.2593             &  & 0.3474             & 0.3904             \\ 
    \cmidrule{3-17}
                                                            &                                                                            & \checkmark                       &            &                                     &            & \checkmark           &  & 0.8344               &  & 0.2386                     &  & 0.2640                       & 0.2752             &  & 0.3634             & 0.4079             \\ 
                                                            &                                                                            &                                  & \checkmark &                                     &            & \checkmark           &  & 0.8627               &  & \underline{0.3027}         &  & \underline{0.3318}           & \underline{0.3432} &  & \underline{0.4425} & 0.4872             \\ 
    \bottomrule
  \end{tabular}%
  \par\smallskip\small
  The best values are \textbf{bolded} and the second best values are \underline{underlined}.

\end{table}

We further investigate the impact of incorporating a graph-based expert, LightGCN, as a long-term preference model within the \proposedcove$_h$ framework. \autoref{tab:w_graph} summarizes the experimental results on both Diginetica and RetailRocket datasets, comparing individual experts as well as \proposedcove$_h$ based on {\bpr} model.

On the Diginetica dataset, integrating LightGCN with short-term experts leads to notable improvements. Specifically, the combination of GRU4Rec and LightGCN achieves the highest AUC score (0.8139), outperforming all other configurations. Additionally, the pairing of SASRec with LightGCN achieves the best Recall@10 (0.4711) and Recall@20 (0.4956). These results demonstrate that LightGCN as the graph-based long-term expert can complement short-term models and enhance overall recommendation.

In contrast, on the RetailRocket dataset, LightGCN as an individual expert does not perform as well as {\bpr}. Consequently, combinations involving LightGCN do not outperform those with BPR in most metrics. However, the combination of {\sasrec} and LightGCN still achieves competitive results, ranking second-best in several Recall and NDCG metrics, despite {\sasrec} and LightGCN performing poorly individually. This suggests that even when individual experts are weak, their combination can provide complementary benefits.

Overall, these findings highlight the potential of integrating graph-based long-term experts, such as LightGCN, within the \proposedcove$_h$ framework, particularly when paired with strong short-term models. This integration can lead to improved recommendation performance, similar to the results achieved with long-term preference models like \bpr.

\subsection{Case Study: Dynamic Gating Mechanism}
To demonstrate the adaptive nature of \proposedcove's gating mechanism, we analyze how the same ground-truth item is recommended to different users through distinct expert weightings.

\begin{figure}[t]
  \centering
  \includegraphics[width=0.5\textwidth]{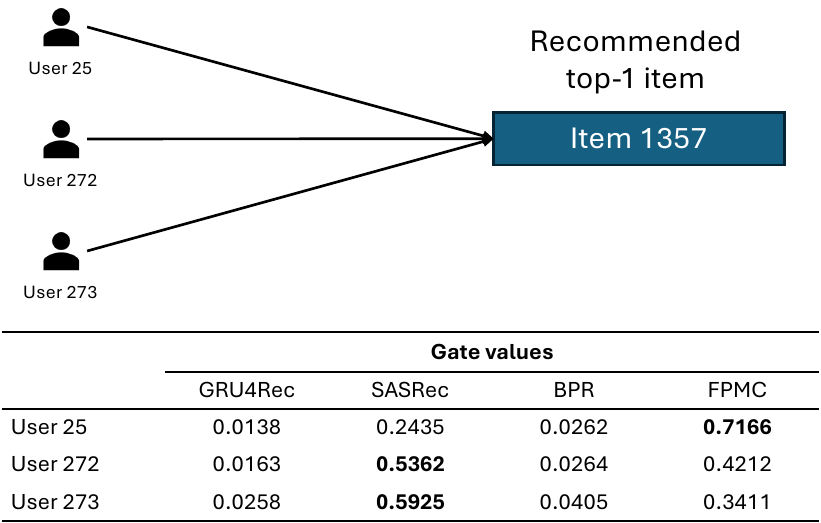}
  \caption{Dynamic gating mechanism: different expert weights for three users being recommended the same item (item 1357). The gating mechanism adapts to each user's unique preference patterns.}%
  \label{fig:case-study-diginetica}%
\end{figure}

\autoref{fig:case-study-diginetica} presents an interesting example from the Diginetica dataset, where item 1357 is correctly recommended to three different users (indices 25, 272, and 273), yet the gating mechanism assigns substantially different expert weights for each user. This demonstrates the dynamic adaptation of \proposedcove\ to individual user contexts.

For user 25, the gating mechanism assigns weights of (0.0138, 0.2435, 0.0262, 0.7166) to GRU4Rec, SASRec, BPR, and FPMC, respectively. Here, FPMC dominates with 71.66\% contribution, indicating that item 1357 aligns with this user's long-term preference patterns. The minimal contributions from GRU4Rec (1.38\%) and BPR (2.62\%) suggest that this recommendation is driven primarily by consistent historical preferences rather than recent interactions.

In contrast, for users 272 and 273, the gating mechanism shifts emphasis toward short-term preference experts. For user 272, the weights are (0.0163, 0.5362, 0.0264, 0.4212), with SASRec contributing 53.62\% and FPMC contributing 42.12\%. For user 273, the weights are (0.0258, 0.5925, 0.0405, 0.3411), with SASRec dominating at 59.25\%. For both users, SASRec emerges as the primary contributor, suggesting that item 1357 matches their recent behavioral patterns and short-term interests.

This case study highlights a key advantage of \proposedcove's dynamic gating mechanism: the same item can be recommended for fundamentally different reasons based on individual user contexts. The gating mechanism automatically identifies whether an item fits a user's long-term stable preferences or their current short-term interests, enabling more nuanced and personalized recommendations. This adaptive behavior would be impossible with static ensemble methods or uniform weighting schemes, validating the importance of learned dynamic gating in mixture-of-experts recommendation systems.



\subsection{LLM as Gating Mechanism}
In this section, we explore the potential of using Large Language Models (LLMs) as a gating mechanism for \proposedcove. The objective is to study if LLMs can dynamically select experts based on user and item information, potentially improving recommendation performance.

We conducted an experiment using ChatGPT-4o\footnote{\url{https://chatgpt.com/share/68511ca8-605c-8012-ac49-2a4aba7f8f66}} to investigate whether LLMs off-the-shelf can effectively aggregate expert predictions. For a given user, we provided the following details: (i) normalized scores from each pretrained expert for all candidate items, (ii) the user's complete historical interaction sequence, and (iii) a request to generate a ranked list of top-50 items. The prompt used is shown below:

\begin{center}
  \fbox{\begin{minipage}{0.95\textwidth}
      \small\ttfamily
      \noindent You are acting as the gating network in a Mixture-of-Experts (MoE) model for sequential recommendation. You will receive inputs corresponding to three users. Each input file is formatted as follows:\\[0.5em]
      \hspace*{1em}Line 1: user\_id\\
      \hspace*{1em}Line 2: A list of sessions (user's interaction history), where each\\
      \hspace*{3em}session is a list of item IDs the user has interacted with.\\
      \hspace*{1em}Lines 3-6: Scores from 4 expert models for the same set of N items.\\
      \hspace*{3em}Each line contains a list of N floats, representing the\\
      \hspace*{3em}predicted relevance scores from that expert for each item.\\[0.5em]
      Your task is to act as the gating network that combines the expert outputs. Please compute a final score for each of the N items per user by appropriately weighting the expert outputs. You should also briefly explain the reasoning behind how the final scores were derived for each user.\\[0.5em]
      For each user, output a CSV file listing the top-50 recommended items, with the following columns: item\_id, final\_score.
    \end{minipage}}
\end{center}

Upon analysis, ChatGPT-4o adopted a simple averaging strategy: it computed the arithmetic mean of scores assigned by each expert and ranked items accordingly. This approach effectively treats all experts equally, similar to a uniform gating mechanism where $g(x)_i = \frac{1}{K}$ for all experts. As the gating mechanism, ChatGPT-4o did not leverage the user's historical interactions to dynamically adjust expert weights or identify user-specific preference patterns.

This simple experiment indicates that, at present, ChatGPT-4o does not provide additional benefits over simple averaging when weighing experts. This suggests that current general-purpose LLMs, when used in a zero-shot manner may not effectively capture the nuanced relationships between user preferences and expert predictions required for optimal gating in recommendation systems.

\subsection{Discussion}\label{sec:discussion}
In this section, we further discuss alternative ways to scale up our proposed \proposedcove\ using a larger model (such as increasing the number of experts) or a larger dataset. In addition, we enumerate a few limitations and potential direction for future improvements.

  {\bf Increasing the number of experts by multiplying the same set of experts.}
Here we try another approach to increase the number of experts in \proposedcove, by multiplying the same set of experts multiple times. As reported in \autoref{tab:multiply-cove-experts}, it is not clear that increasing the number of the same experts multiple times helps in enhancing recommendation performance. However, this approach also increases the computational cost of training due to the larger number of experts being used.

\begin{table}[t]
  \small
  \centering
  \caption{Performance when increasing the number of experts: AUC and MRR}
  \label{tab:multiply-cove-experts}
  \setlength{\tabcolsep}{5pt}
  \begin{tabular}{ccccccccccccccccc}
    \toprule
    \multirow{4}{*}{Multiply} &  & \multicolumn{5}{c}{Diginetica}        &        & \multicolumn{5}{c}{RetailRocket}                                                                                                                                \\
    \cmidrule{3-7}
    \cmidrule{9-13}
                              &  & \multicolumn{2}{c}{\proposedcove$_h$} &        & \multicolumn{2}{c}{\proposedcove$_s$} &        & \multicolumn{2}{c}{\proposedcove$_h$} &  & \multicolumn{2}{c}{\proposedcove$_s$}                               \\
    \cmidrule{3-4}
    \cmidrule{6-7}
    \cmidrule{9-10}
    \cmidrule{12-13}
                              &  & AUC                                   & MRR    &                                       & AUC    & MRR                                   &  & AUC                                   & MRR    &  & AUC    & MRR    \\
    \midrule
    $\times1$                 &  & 0.8007                                & 0.3756 &                                       & 0.7854 & 0.4112                                &  & 0.8798                                & 0.3406 &  & 0.8767 & 0.3365 \\
    $\times2$                 &  & 0.7982                                & 0.3731 &                                       & 0.7964 & 0.4097                                &  & 0.8826                                & 0.3419 &  & 0.8755 & 0.3374 \\
    $\times4$                 &  & 0.7984                                & 0.3842 &                                       & 0.7963 & 0.4125                                &  & 0.8797                                & 0.3406 &  & 0.8738 & 0.3383 \\
    \bottomrule
  \end{tabular}
\end{table}

%

{\bf CoVE$\boldsymbol{_h}$ and CoVE$\boldsymbol{_s}$ choices.} The two proposed \proposedcove\ frameworks offer different levels of combining experts. \proposedcove$_h$ combines experts at the hidden layer, allowing for more complex interactions between expert outputs. This also limits the particular experts that can be used, as they must have the same output hidden dimension.  In contrast, \proposedcove$_s$ combines experts at the output layer, which is simpler and more interpretable. We can utilize any experts with different output hidden dimensions, as long as they can output a score for the set of items. The gating mechanism in \proposedcove$_s$ is also simpler, as it only needs to combine the output scores of the experts rather than their hidden states. However, this simplicity comes at the cost of potentially losing some complex interactions between expert outputs that \proposedcove$_h$ can capture.
The choice between these two variants depends on the specific use case and the desired complexity of the model. For example, \proposedcove$_h$ may be more suitable for tasks requiring complex interactions between expert outputs, while \proposedcove$_s$ may be preferred for tasks where interpretability is crucial.

  {\bf Running time analysis.}
\proposedcove\ models are mixtures of all individual experts, so the overall training time is approximately the sum of all experts' training time, overhead from the gating mechanism, and the time saved by requiring only a single backward pass for the entire model instead of separate backward passes for individual experts.

~\autoref{tab:running-time} reports the training time for the two \proposedcove\ variants that incorporate all four experts (\bpr, \fpmc, \gruforrec, and \sasrec), along with the training time for each expert on Diginetica dataset. All models were trained for $200$ epochs with a batch size of $32$, and running time was measured in seconds. Among the individual experts, \bpr\ required the least time to train ($7.5$ minutes), while \fpmc\ required the most ($19.1$ minutes). \gruforrec\ and \sasrec\ took similar training times ($11.3$ and $11.8$ minutes, respectively). The total training time for all four experts was $49.6$ minutes. In comparison, \proposedcove$_h$ took $78$ minutes, and \proposedcove$_s$ took $62.8$ minutes, corresponding to approximately $1.6$ and $1.3$ times the combined training time of the individual experts, respectively. This increase is expected, as \proposedcove\ models incur additional computational costs from the gating mechanism and the aggregation of expert outputs.

In practice, the use of pretrained individual experts can substantially reduce the training time for \proposedcove\ variants. When pretrained experts are available, they can be directly incorporated into \proposedcove\ without training from scratch, enabling much faster convergence of the \proposedcove\ models.

\begin{table}[]
  \small
  \caption{Training time for Diginetica dataset, same batch size (32) and number of epochs (200) for all models.}
  \label{tab:running-time}
  \begin{tabular}{c|cccc|c|cc}
    \toprule
    Model             & \bpr    & \fpmc    & \gruforrec & \sasrec & Expert total & \proposedcove$_h$ & \proposedcove$_s$ \\
    \midrule
    Training time (s) & 449.256 & 1146.133 & 676.165    & 705.373 & 2976.927     & 4677.913          & 3770.832          \\
    \bottomrule
  \end{tabular}
\end{table}

{\bf Limitations and potential improvements}. Although both of \proposedcove\ variants achieve consistently better performance, some limitations persist in the chosen approaches. Firstly, \proposedcove\ uses continuous mixture-of-experts training, where every single expert is activated. This may result in a longer and inefficient training process and difficulties in scalability. However, due to the uniqueness of how each expert views input data, continuous training seems necessary for the experts to capture complete user behaviors.  Secondly, various experts may converge differently, which the proposed methods did not tackle and could lead to overfitting for individual experts when the overall objective is achieved.  The Composition of Variant Experts framework is generalized and flexible in incorporating any type of experts; however, in this work, expert selection mainly focuses on a few well-known approaches.
Furthermore, we presuppose the output hidden dimension of each expert in the proposed \proposedcove$_h$ variant to be the same. This limits the flexibility of each individual expert, as the best-performing parameters for one expert may differ from another in practice. Methods of combining experts with different hidden output dimensions could be considered.

\section{Conclusion}

While users typically have a long-term preference profile built up over a long period of time, there are short-term effects that may affect preferences as well.  In this work, we establish that different types of `experts' that focus on either short- or long-term preference would work better together, as not only do both types of preferences matter, but the way they do so differs from user to user.  Which model predominates also varies across datasets, emphasizing the proposed framework approach of \proposedcove, leveraging on multiple experts working in concert, while accommodating a gating mechanism that modulates the influence of each expert on recommendations.



\bibliographystyle{ACM-Reference-Format}
\bibliography{ref}


\begin{thebibliography}{78}


\ifx \showCODEN    \undefined \def \showCODEN     #1{\unskip}     \fi
\ifx \showISBNx    \undefined \def \showISBNx     #1{\unskip}     \fi
\ifx \showISBNxiii \undefined \def \showISBNxiii  #1{\unskip}     \fi
\ifx \showISSN     \undefined \def \showISSN      #1{\unskip}     \fi
\ifx \showLCCN     \undefined \def \showLCCN      #1{\unskip}     \fi
\ifx \shownote     \undefined \def \shownote      #1{#1}          \fi
\ifx \showarticletitle \undefined \def \showarticletitle #1{#1}   \fi
\ifx \showURL      \undefined \def \showURL       {\relax}        \fi
\providecommand\bibfield[2]{#2}
\providecommand\bibinfo[2]{#2}
\providecommand\natexlab[1]{#1}
\providecommand\showeprint[2][]{arXiv:#2}

\bibitem[Abbattista et~al\mbox{.}(2024)]%
        {abbattista2024enhancing}
\bibfield{author}{\bibinfo{person}{Davide Abbattista}, \bibinfo{person}{Vito~Walter Anelli}, \bibinfo{person}{Tommaso Di~Noia}, \bibinfo{person}{Craig Macdonald}, {and} \bibinfo{person}{Aleksandr~Vladimirovich Petrov}.} \bibinfo{year}{2024}\natexlab{}.
\newblock \showarticletitle{Enhancing Sequential Music Recommendation with Personalized Popularity Awareness}. In \bibinfo{booktitle}{\emph{Proceedings of the 18th ACM Conference on Recommender Systems}}. \bibinfo{pages}{1168--1173}.
\newblock


\bibitem[Bhagat et~al\mbox{.}(2018)]%
        {bhagat2018buy}
\bibfield{author}{\bibinfo{person}{Rahul Bhagat}, \bibinfo{person}{Srevatsan Muralidharan}, \bibinfo{person}{Alex Lobzhanidze}, {and} \bibinfo{person}{Shankar Vishwanath}.} \bibinfo{year}{2018}\natexlab{}.
\newblock \showarticletitle{Buy it again: Modeling repeat purchase recommendations}. In \bibinfo{booktitle}{\emph{Proceedings of the 24th ACM SIGKDD international conference on knowledge discovery \& data mining}}. \bibinfo{pages}{62--70}.
\newblock


\bibitem[Bian et~al\mbox{.}(2023)]%
        {bian2023multimodal}
\bibfield{author}{\bibinfo{person}{Shuqing Bian}, \bibinfo{person}{Xingyu Pan}, \bibinfo{person}{Wayne~Xin Zhao}, \bibinfo{person}{Jinpeng Wang}, \bibinfo{person}{Chuyuan Wang}, {and} \bibinfo{person}{Ji-Rong Wen}.} \bibinfo{year}{2023}\natexlab{}.
\newblock \showarticletitle{Multi-modal Mixture of Experts Represetation Learning for Sequential Recommendation}. In \bibinfo{booktitle}{\emph{Proceedings of the 32nd ACM International Conference on Information and Knowledge Management}} (Birmingham, United Kingdom) \emph{(\bibinfo{series}{CIKM '23})}. \bibinfo{publisher}{Association for Computing Machinery}, \bibinfo{address}{New York, NY, USA}, \bibinfo{pages}{110–119}.
\newblock
\showISBNx{9798400701245}
\href{https://doi.org/10.1145/3583780.3614978}{doi:\nolinkurl{10.1145/3583780.3614978}}


\bibitem[Ca{\~n}amares and Castells(2020)]%
        {canamares2020target}
\bibfield{author}{\bibinfo{person}{Roc{\'\i}o Ca{\~n}amares} {and} \bibinfo{person}{Pablo Castells}.} \bibinfo{year}{2020}\natexlab{}.
\newblock \showarticletitle{On target item sampling in offline recommender system evaluation}. In \bibinfo{booktitle}{\emph{Proceedings of the 14th ACM Conference on Recommender Systems}}. \bibinfo{pages}{259--268}.
\newblock


\bibitem[Chen et~al\mbox{.}(2025)]%
        {chen2025enhancing}
\bibfield{author}{\bibinfo{person}{Lei Chen}, \bibinfo{person}{Chen Gao}, \bibinfo{person}{Xiaoyi Du}, \bibinfo{person}{Hengliang Luo}, \bibinfo{person}{Depeng Jin}, \bibinfo{person}{Yong Li}, {and} \bibinfo{person}{Meng Wang}.} \bibinfo{year}{2025}\natexlab{}.
\newblock \showarticletitle{Enhancing id-based recommendation with large language models}.
\newblock \bibinfo{journal}{\emph{ACM Transactions on Information Systems}} \bibinfo{volume}{43}, \bibinfo{number}{5} (\bibinfo{year}{2025}), \bibinfo{pages}{1--30}.
\newblock


\bibitem[Chen et~al\mbox{.}(2022)]%
        {chen2022double}
\bibfield{author}{\bibinfo{person}{Qi Chen}, \bibinfo{person}{Guohui Li}, \bibinfo{person}{Quan Zhou}, \bibinfo{person}{Si Shi}, {and} \bibinfo{person}{Deqing Zou}.} \bibinfo{year}{2022}\natexlab{}.
\newblock \showarticletitle{Double attention convolutional neural network for sequential recommendation}.
\newblock \bibinfo{journal}{\emph{ACM Transactions on the Web}} \bibinfo{volume}{16}, \bibinfo{number}{4} (\bibinfo{year}{2022}), \bibinfo{pages}{1--23}.
\newblock


\bibitem[Chen et~al\mbox{.}(2024)]%
        {chen2024survey}
\bibfield{author}{\bibinfo{person}{Shu Chen}, \bibinfo{person}{Zitao Xu}, \bibinfo{person}{Weike Pan}, \bibinfo{person}{Qiang Yang}, {and} \bibinfo{person}{Zhong Ming}.} \bibinfo{year}{2024}\natexlab{}.
\newblock \showarticletitle{A survey on cross-domain sequential recommendation}. In \bibinfo{booktitle}{\emph{Proceedings of the Thirty-Third International Joint Conference on Artificial Intelligence}}. \bibinfo{pages}{7989--7998}.
\newblock


\bibitem[Chen et~al\mbox{.}(2019)]%
        {chen2019personalized}
\bibfield{author}{\bibinfo{person}{Xu Chen}, \bibinfo{person}{Hanxiong Chen}, \bibinfo{person}{Hongteng Xu}, \bibinfo{person}{Yongfeng Zhang}, \bibinfo{person}{Yixin Cao}, \bibinfo{person}{Zheng Qin}, {and} \bibinfo{person}{Hongyuan Zha}.} \bibinfo{year}{2019}\natexlab{}.
\newblock \showarticletitle{Personalized Fashion Recommendation with Visual Explanations based on Multimodal Attention Network: Towards Visually Explainable Recommendation}. In \bibinfo{booktitle}{\emph{Proceedings of the 42nd International ACM SIGIR Conference on Research and Development in Information Retrieval}} (Paris, France) \emph{(\bibinfo{series}{SIGIR'19})}. \bibinfo{publisher}{Association for Computing Machinery}, \bibinfo{address}{New York, NY, USA}, \bibinfo{pages}{765–774}.
\newblock
\showISBNx{9781450361729}
\href{https://doi.org/10.1145/3331184.3331254}{doi:\nolinkurl{10.1145/3331184.3331254}}


\bibitem[Chen et~al\mbox{.}(2023)]%
        {chen2023survey}
\bibfield{author}{\bibinfo{person}{Xiaoqing Chen}, \bibinfo{person}{Zhitao Li}, \bibinfo{person}{Weike Pan}, {and} \bibinfo{person}{Zhong Ming}.} \bibinfo{year}{2023}\natexlab{}.
\newblock \showarticletitle{A survey on multi-behavior sequential recommendation}.
\newblock \bibinfo{journal}{\emph{arXiv preprint arXiv:2308.15701}} (\bibinfo{year}{2023}).
\newblock


\bibitem[Chung et~al\mbox{.}(2014)]%
        {chung2014empirical}
\bibfield{author}{\bibinfo{person}{Junyoung Chung}, \bibinfo{person}{Caglar Gulcehre}, \bibinfo{person}{KyungHyun Cho}, {and} \bibinfo{person}{Yoshua Bengio}.} \bibinfo{year}{2014}\natexlab{}.
\newblock \showarticletitle{Empirical evaluation of gated recurrent neural networks on sequence modeling}.
\newblock \bibinfo{journal}{\emph{arXiv preprint arXiv:1412.3555}} (\bibinfo{year}{2014}).
\newblock


\bibitem[Clark et~al\mbox{.}(2022)]%
        {clark2022unified}
\bibfield{author}{\bibinfo{person}{Aidan Clark}, \bibinfo{person}{Diego de Las~Casas}, \bibinfo{person}{Aurelia Guy}, \bibinfo{person}{Arthur Mensch}, \bibinfo{person}{Michela Paganini}, \bibinfo{person}{Jordan Hoffmann}, \bibinfo{person}{Bogdan Damoc}, \bibinfo{person}{Blake Hechtman}, \bibinfo{person}{Trevor Cai}, \bibinfo{person}{Sebastian Borgeaud}, {et~al\mbox{.}}} \bibinfo{year}{2022}\natexlab{}.
\newblock \showarticletitle{Unified scaling laws for routed language models}. In \bibinfo{booktitle}{\emph{International conference on machine learning}}. PMLR, \bibinfo{pages}{4057--4086}.
\newblock


\bibitem[Dallmann et~al\mbox{.}(2021)]%
        {dallmann2021case}
\bibfield{author}{\bibinfo{person}{Alexander Dallmann}, \bibinfo{person}{Daniel Zoller}, {and} \bibinfo{person}{Andreas Hotho}.} \bibinfo{year}{2021}\natexlab{}.
\newblock \showarticletitle{A case study on sampling strategies for evaluating neural sequential item recommendation models}. In \bibinfo{booktitle}{\emph{Proceedings of the 15th ACM Conference on Recommender Systems}}. \bibinfo{pages}{505--514}.
\newblock


\bibitem[de~Souza Pereira~Moreira et~al\mbox{.}(2021)]%
        {de2021transformers4rec}
\bibfield{author}{\bibinfo{person}{Gabriel de Souza Pereira~Moreira}, \bibinfo{person}{Sara Rabhi}, \bibinfo{person}{Jeong~Min Lee}, \bibinfo{person}{Ronay Ak}, {and} \bibinfo{person}{Even Oldridge}.} \bibinfo{year}{2021}\natexlab{}.
\newblock \showarticletitle{Transformers4rec: Bridging the gap between nlp and sequential/session-based recommendation}. In \bibinfo{booktitle}{\emph{Proceedings of the 15th ACM conference on recommender systems}}. \bibinfo{pages}{143--153}.
\newblock


\bibitem[Do and Lauw(2023)]%
        {do2023continual}
\bibfield{author}{\bibinfo{person}{Jaime~Hieu Do} {and} \bibinfo{person}{Hady~W Lauw}.} \bibinfo{year}{2023}\natexlab{}.
\newblock \showarticletitle{Continual collaborative filtering through gradient alignment}. In \bibinfo{booktitle}{\emph{Proceedings of the 17th ACM Conference on Recommender Systems}}. \bibinfo{pages}{1133--1138}.
\newblock


\bibitem[Geng et~al\mbox{.}(2022)]%
        {geng2022recommendation}
\bibfield{author}{\bibinfo{person}{Shijie Geng}, \bibinfo{person}{Shuchang Liu}, \bibinfo{person}{Zuohui Fu}, \bibinfo{person}{Yingqiang Ge}, {and} \bibinfo{person}{Yongfeng Zhang}.} \bibinfo{year}{2022}\natexlab{}.
\newblock \showarticletitle{Recommendation as language processing (rlp): A unified pretrain, personalized prompt \& predict paradigm (p5)}. In \bibinfo{booktitle}{\emph{Proceedings of the 16th ACM conference on recommender systems}}. \bibinfo{pages}{299--315}.
\newblock


\bibitem[Hazimeh et~al\mbox{.}(2021)]%
        {hazimeh2021dselect}
\bibfield{author}{\bibinfo{person}{Hussein Hazimeh}, \bibinfo{person}{Zhe Zhao}, \bibinfo{person}{Aakanksha Chowdhery}, \bibinfo{person}{Maheswaran Sathiamoorthy}, \bibinfo{person}{Yihua Chen}, \bibinfo{person}{Rahul Mazumder}, \bibinfo{person}{Lichan Hong}, {and} \bibinfo{person}{Ed Chi}.} \bibinfo{year}{2021}\natexlab{}.
\newblock \showarticletitle{Dselect-k: Differentiable selection in the mixture of experts with applications to multi-task learning}.
\newblock \bibinfo{journal}{\emph{Advances in Neural Information Processing Systems}}  \bibinfo{volume}{34} (\bibinfo{year}{2021}), \bibinfo{pages}{29335--29347}.
\newblock


\bibitem[He et~al\mbox{.}(2017)]%
        {he2017translation}
\bibfield{author}{\bibinfo{person}{Ruining He}, \bibinfo{person}{Wang-Cheng Kang}, {and} \bibinfo{person}{Julian McAuley}.} \bibinfo{year}{2017}\natexlab{}.
\newblock \showarticletitle{Translation-based Recommendation}. In \bibinfo{booktitle}{\emph{Proceedings of the Eleventh ACM Conference on Recommender Systems}} (Como, Italy) \emph{(\bibinfo{series}{RecSys '17})}. \bibinfo{publisher}{Association for Computing Machinery}, \bibinfo{address}{New York, NY, USA}, \bibinfo{pages}{161–169}.
\newblock
\showISBNx{9781450346528}
\href{https://doi.org/10.1145/3109859.3109882}{doi:\nolinkurl{10.1145/3109859.3109882}}


\bibitem[He et~al\mbox{.}(2015)]%
        {he2015trirank}
\bibfield{author}{\bibinfo{person}{Xiangnan He}, \bibinfo{person}{Tao Chen}, \bibinfo{person}{Min-Yen Kan}, {and} \bibinfo{person}{Xiao Chen}.} \bibinfo{year}{2015}\natexlab{}.
\newblock \showarticletitle{TriRank: Review-aware Explainable Recommendation by Modeling Aspects}. In \bibinfo{booktitle}{\emph{Proceedings of the 24th ACM International on Conference on Information and Knowledge Management}} (Melbourne, Australia) \emph{(\bibinfo{series}{CIKM '15})}. \bibinfo{publisher}{Association for Computing Machinery}, \bibinfo{address}{New York, NY, USA}, \bibinfo{pages}{1661–1670}.
\newblock
\showISBNx{9781450337946}
\href{https://doi.org/10.1145/2806416.2806504}{doi:\nolinkurl{10.1145/2806416.2806504}}


\bibitem[He et~al\mbox{.}(2020)]%
        {he2020lightgcn}
\bibfield{author}{\bibinfo{person}{Xiangnan He}, \bibinfo{person}{Kuan Deng}, \bibinfo{person}{Xiang Wang}, \bibinfo{person}{Yan Li}, \bibinfo{person}{Yongdong Zhang}, {and} \bibinfo{person}{Meng Wang}.} \bibinfo{year}{2020}\natexlab{}.
\newblock \showarticletitle{Lightgcn: Simplifying and powering graph convolution network for recommendation}. In \bibinfo{booktitle}{\emph{Proceedings of the 43rd International ACM SIGIR conference on research and development in Information Retrieval}}. \bibinfo{pages}{639--648}.
\newblock


\bibitem[He et~al\mbox{.}(2016)]%
        {he2016fast}
\bibfield{author}{\bibinfo{person}{Xiangnan He}, \bibinfo{person}{Hanwang Zhang}, \bibinfo{person}{Min-Yen Kan}, {and} \bibinfo{person}{Tat-Seng Chua}.} \bibinfo{year}{2016}\natexlab{}.
\newblock \showarticletitle{Fast matrix factorization for online recommendation with implicit feedback}. In \bibinfo{booktitle}{\emph{Proceedings of the 39th International ACM SIGIR conference on Research and Development in Information Retrieval}}. \bibinfo{pages}{549--558}.
\newblock


\bibitem[Hidasi and Czapp(2023a)]%
        {hidasi2023effect}
\bibfield{author}{\bibinfo{person}{Bal{\'a}zs Hidasi} {and} \bibinfo{person}{{\'A}d{\'a}m~Tibor Czapp}.} \bibinfo{year}{2023}\natexlab{a}.
\newblock \showarticletitle{The effect of third party implementations on reproducibility}. In \bibinfo{booktitle}{\emph{Proceedings of the 17th ACM Conference on Recommender Systems}}. \bibinfo{pages}{272--282}.
\newblock


\bibitem[Hidasi and Czapp(2023b)]%
        {hidasi2023widespread}
\bibfield{author}{\bibinfo{person}{Bal{\'a}zs Hidasi} {and} \bibinfo{person}{{\'A}d{\'a}m~Tibor Czapp}.} \bibinfo{year}{2023}\natexlab{b}.
\newblock \showarticletitle{Widespread flaws in offline evaluation of recommender systems}. In \bibinfo{booktitle}{\emph{Proceedings of the 17th acm conference on recommender systems}}. \bibinfo{pages}{848--855}.
\newblock


\bibitem[Hidasi and Karatzoglou(2018)]%
        {hidasi2018recurrent}
\bibfield{author}{\bibinfo{person}{Bal{\'a}zs Hidasi} {and} \bibinfo{person}{Alexandros Karatzoglou}.} \bibinfo{year}{2018}\natexlab{}.
\newblock \showarticletitle{Recurrent neural networks with top-k gains for session-based recommendations}. In \bibinfo{booktitle}{\emph{Proceedings of the 27th ACM international conference on information and knowledge management}}. \bibinfo{pages}{843--852}.
\newblock


\bibitem[Hidasi et~al\mbox{.}(2016)]%
        {gru4rec1}
\bibfield{author}{\bibinfo{person}{Bal{\'{a}}zs Hidasi}, \bibinfo{person}{Alexandros Karatzoglou}, \bibinfo{person}{Linas Baltrunas}, {and} \bibinfo{person}{Domonkos Tikk}.} \bibinfo{year}{2016}\natexlab{}.
\newblock \showarticletitle{Session-based Recommendations with Recurrent Neural Networks}. In \bibinfo{booktitle}{\emph{4th International Conference on Learning Representations, {ICLR} 2016, San Juan, Puerto Rico, May 2-4, 2016, Conference Track Proceedings}}.
\newblock
\urldef\tempurl%
\url{http://arxiv.org/abs/1511.06939}
\showURL{%
\tempurl}


\bibitem[Hou et~al\mbox{.}(2023)]%
        {hou2023learning}
\bibfield{author}{\bibinfo{person}{Yupeng Hou}, \bibinfo{person}{Zhankui He}, \bibinfo{person}{Julian McAuley}, {and} \bibinfo{person}{Wayne~Xin Zhao}.} \bibinfo{year}{2023}\natexlab{}.
\newblock \showarticletitle{Learning vector-quantized item representation for transferable sequential recommenders}. In \bibinfo{booktitle}{\emph{Proceedings of the ACM Web Conference 2023}}. \bibinfo{pages}{1162--1171}.
\newblock


\bibitem[Hu et~al\mbox{.}(2008)]%
        {hu2008collaborative}
\bibfield{author}{\bibinfo{person}{Yifan Hu}, \bibinfo{person}{Yehuda Koren}, {and} \bibinfo{person}{Chris Volinsky}.} \bibinfo{year}{2008}\natexlab{}.
\newblock \showarticletitle{Collaborative filtering for implicit feedback datasets}. In \bibinfo{booktitle}{\emph{2008 Eighth IEEE international conference on data mining}}. Ieee, \bibinfo{pages}{263--272}.
\newblock


\bibitem[Jacobs et~al\mbox{.}(1991)]%
        {jacobs1991adaptive}
\bibfield{author}{\bibinfo{person}{Robert~A Jacobs}, \bibinfo{person}{Michael~I Jordan}, \bibinfo{person}{Steven~J Nowlan}, {and} \bibinfo{person}{Geoffrey~E Hinton}.} \bibinfo{year}{1991}\natexlab{}.
\newblock \showarticletitle{Adaptive mixtures of local experts}.
\newblock \bibinfo{journal}{\emph{Neural computation}} \bibinfo{volume}{3}, \bibinfo{number}{1} (\bibinfo{year}{1991}), \bibinfo{pages}{79--87}.
\newblock


\bibitem[Jendal et~al\mbox{.}(2024)]%
        {jendal2024hypergraphs}
\bibfield{author}{\bibinfo{person}{Theis~E. Jendal}, \bibinfo{person}{Trung-Hoang Le}, \bibinfo{person}{Hady~W. Lauw}, \bibinfo{person}{Matteo Lissandrini}, \bibinfo{person}{Peter Dolog}, {and} \bibinfo{person}{Katja Hose}.} \bibinfo{year}{2024}\natexlab{}.
\newblock \showarticletitle{Hypergraphs with Attention on Reviews for Explainable Recommendation}. In \bibinfo{booktitle}{\emph{Advances in Information Retrieval}}, \bibfield{editor}{\bibinfo{person}{Nazli Goharian}, \bibinfo{person}{Nicola Tonellotto}, \bibinfo{person}{Yulan He}, \bibinfo{person}{Aldo Lipani}, \bibinfo{person}{Graham McDonald}, \bibinfo{person}{Craig Macdonald}, {and} \bibinfo{person}{Iadh Ounis}} (Eds.). \bibinfo{publisher}{Springer Nature Switzerland}, \bibinfo{address}{Cham}, \bibinfo{pages}{230--246}.
\newblock
\showISBNx{978-3-031-56027-9}


\bibitem[Jiang et~al\mbox{.}(2024)]%
        {jiang2024mixtral}
\bibfield{author}{\bibinfo{person}{Albert~Q Jiang}, \bibinfo{person}{Alexandre Sablayrolles}, \bibinfo{person}{Antoine Roux}, \bibinfo{person}{Arthur Mensch}, \bibinfo{person}{Blanche Savary}, \bibinfo{person}{Chris Bamford}, \bibinfo{person}{Devendra~Singh Chaplot}, \bibinfo{person}{Diego de~las Casas}, \bibinfo{person}{Emma~Bou Hanna}, \bibinfo{person}{Florian Bressand}, {et~al\mbox{.}}} \bibinfo{year}{2024}\natexlab{}.
\newblock \showarticletitle{Mixtral of experts}.
\newblock \bibinfo{journal}{\emph{arXiv preprint arXiv:2401.04088}} (\bibinfo{year}{2024}).
\newblock


\bibitem[Kang and McAuley(2018)]%
        {kang2018self}
\bibfield{author}{\bibinfo{person}{Wang-Cheng Kang} {and} \bibinfo{person}{Julian McAuley}.} \bibinfo{year}{2018}\natexlab{}.
\newblock \showarticletitle{Self-attentive sequential recommendation}. In \bibinfo{booktitle}{\emph{2018 IEEE international conference on data mining (ICDM)}}. IEEE, \bibinfo{pages}{197--206}.
\newblock


\bibitem[Kim et~al\mbox{.}(2025)]%
        {kim2025lost}
\bibfield{author}{\bibinfo{person}{Sein Kim}, \bibinfo{person}{Hongseok Kang}, \bibinfo{person}{Kibum Kim}, \bibinfo{person}{Jiwan Kim}, \bibinfo{person}{Donghyun Kim}, \bibinfo{person}{Minchul Yang}, \bibinfo{person}{Kwangjin Oh}, \bibinfo{person}{Julian McAuley}, {and} \bibinfo{person}{Chanyoung Park}.} \bibinfo{year}{2025}\natexlab{}.
\newblock \showarticletitle{Lost in Sequence: Do Large Language Models Understand Sequential Recommendation?}. In \bibinfo{booktitle}{\emph{Proceedings of the 31st ACM SIGKDD Conference on Knowledge Discovery and Data Mining V. 2}}. \bibinfo{pages}{1160--1171}.
\newblock


\bibitem[Klenitskiy and Vasilev(2023)]%
        {klenitskiy2023turning}
\bibfield{author}{\bibinfo{person}{Anton Klenitskiy} {and} \bibinfo{person}{Alexey Vasilev}.} \bibinfo{year}{2023}\natexlab{}.
\newblock \showarticletitle{Turning Dross Into Gold Loss: is BERT4Rec really better than SASRec?}. In \bibinfo{booktitle}{\emph{Proceedings of the 17th ACM Conference on Recommender Systems}}. \bibinfo{pages}{1120--1125}.
\newblock


\bibitem[Koren et~al\mbox{.}(2009)]%
        {koren2009matrix}
\bibfield{author}{\bibinfo{person}{Yehuda Koren}, \bibinfo{person}{Robert Bell}, {and} \bibinfo{person}{Chris Volinsky}.} \bibinfo{year}{2009}\natexlab{}.
\newblock \showarticletitle{Matrix factorization techniques for recommender systems}.
\newblock \bibinfo{journal}{\emph{Computer}} \bibinfo{volume}{42}, \bibinfo{number}{8} (\bibinfo{year}{2009}), \bibinfo{pages}{30--37}.
\newblock


\bibitem[Krichene and Rendle(2020)]%
        {krichene2020sampled}
\bibfield{author}{\bibinfo{person}{Walid Krichene} {and} \bibinfo{person}{Steffen Rendle}.} \bibinfo{year}{2020}\natexlab{}.
\newblock \showarticletitle{On sampled metrics for item recommendation}. In \bibinfo{booktitle}{\emph{Proceedings of the 26th ACM SIGKDD international conference on knowledge discovery \& data mining}}. \bibinfo{pages}{1748--1757}.
\newblock


\bibitem[Latifi et~al\mbox{.}(2021)]%
        {latifi2021session}
\bibfield{author}{\bibinfo{person}{Sara Latifi}, \bibinfo{person}{Noemi Mauro}, {and} \bibinfo{person}{Dietmar Jannach}.} \bibinfo{year}{2021}\natexlab{}.
\newblock \showarticletitle{Session-aware recommendation: A surprising quest for the state-of-the-art}.
\newblock \bibinfo{journal}{\emph{Information Sciences}}  \bibinfo{volume}{573} (\bibinfo{year}{2021}), \bibinfo{pages}{291--315}.
\newblock


\bibitem[Le and Lauw(2024)]%
        {le2024question}
\bibfield{author}{\bibinfo{person}{Trung-Hoang Le} {and} \bibinfo{person}{Hady~W. Lauw}.} \bibinfo{year}{2024}\natexlab{}.
\newblock \showarticletitle{Question-Attentive Review-Level Explanation for Neural Rating Regression}.
\newblock \bibinfo{journal}{\emph{ACM Trans. Intell. Syst. Technol.}} \bibinfo{volume}{15}, \bibinfo{number}{6}, Article \bibinfo{articleno}{132} (\bibinfo{date}{Dec.} \bibinfo{year}{2024}), \bibinfo{numpages}{25}~pages.
\newblock
\showISSN{2157-6904}
\href{https://doi.org/10.1145/3699516}{doi:\nolinkurl{10.1145/3699516}}


\bibitem[Li et~al\mbox{.}(2017)]%
        {li2017neural}
\bibfield{author}{\bibinfo{person}{Jing Li}, \bibinfo{person}{Pengjie Ren}, \bibinfo{person}{Zhumin Chen}, \bibinfo{person}{Zhaochun Ren}, \bibinfo{person}{Tao Lian}, {and} \bibinfo{person}{Jun Ma}.} \bibinfo{year}{2017}\natexlab{}.
\newblock \showarticletitle{Neural attentive session-based recommendation}. In \bibinfo{booktitle}{\emph{Proceedings of the 2017 ACM on Conference on Information and Knowledge Management}}. \bibinfo{pages}{1419--1428}.
\newblock


\bibitem[Li et~al\mbox{.}(2023)]%
        {li2023exploring}
\bibfield{author}{\bibinfo{person}{Ruyu Li}, \bibinfo{person}{Wenhao Deng}, \bibinfo{person}{Yu Cheng}, \bibinfo{person}{Zheng Yuan}, \bibinfo{person}{Jiaqi Zhang}, {and} \bibinfo{person}{Fajie Yuan}.} \bibinfo{year}{2023}\natexlab{}.
\newblock \showarticletitle{Exploring the upper limits of text-based collaborative filtering using large language models: Discoveries and insights}.
\newblock \bibinfo{journal}{\emph{arXiv preprint arXiv:2305.11700}} (\bibinfo{year}{2023}).
\newblock


\bibitem[Li et~al\mbox{.}(2024)]%
        {li2024graph}
\bibfield{author}{\bibinfo{person}{Zihao Li}, \bibinfo{person}{Chao Yang}, \bibinfo{person}{Yakun Chen}, \bibinfo{person}{Xianzhi Wang}, \bibinfo{person}{Hongxu Chen}, \bibinfo{person}{Guandong Xu}, \bibinfo{person}{Lina Yao}, {and} \bibinfo{person}{Michael Sheng}.} \bibinfo{year}{2024}\natexlab{}.
\newblock \showarticletitle{Graph and sequential neural networks in session-based recommendation: A survey}.
\newblock \bibinfo{journal}{\emph{Comput. Surveys}} \bibinfo{volume}{57}, \bibinfo{number}{2} (\bibinfo{year}{2024}), \bibinfo{pages}{1--37}.
\newblock


\bibitem[Liu et~al\mbox{.}(2018)]%
        {liu2018stamp}
\bibfield{author}{\bibinfo{person}{Qiao Liu}, \bibinfo{person}{Yifu Zeng}, \bibinfo{person}{Refuoe Mokhosi}, {and} \bibinfo{person}{Haibin Zhang}.} \bibinfo{year}{2018}\natexlab{}.
\newblock \showarticletitle{STAMP: short-term attention/memory priority model for session-based recommendation}. In \bibinfo{booktitle}{\emph{Proceedings of the 24th ACM SIGKDD international conference on knowledge discovery \& data mining}}. \bibinfo{pages}{1831--1839}.
\newblock


\bibitem[Ludewig et~al\mbox{.}(2019)]%
        {ludewig2019performance}
\bibfield{author}{\bibinfo{person}{Malte Ludewig}, \bibinfo{person}{Noemi Mauro}, \bibinfo{person}{Sara Latifi}, {and} \bibinfo{person}{Dietmar Jannach}.} \bibinfo{year}{2019}\natexlab{}.
\newblock \showarticletitle{Performance comparison of neural and non-neural approaches to session-based recommendation}. In \bibinfo{booktitle}{\emph{Proceedings of the 13th ACM conference on recommender systems}}. \bibinfo{pages}{462--466}.
\newblock


\bibitem[Ma et~al\mbox{.}(2008)]%
        {ma2008sorec}
\bibfield{author}{\bibinfo{person}{Hao Ma}, \bibinfo{person}{Haixuan Yang}, \bibinfo{person}{Michael~R. Lyu}, {and} \bibinfo{person}{Irwin King}.} \bibinfo{year}{2008}\natexlab{}.
\newblock \showarticletitle{SoRec: social recommendation using probabilistic matrix factorization}. In \bibinfo{booktitle}{\emph{Proceedings of the 17th ACM Conference on Information and Knowledge Management}} (Napa Valley, California, USA) \emph{(\bibinfo{series}{CIKM '08})}. \bibinfo{publisher}{Association for Computing Machinery}, \bibinfo{address}{New York, NY, USA}, \bibinfo{pages}{931–940}.
\newblock
\showISBNx{9781595939913}
\href{https://doi.org/10.1145/1458082.1458205}{doi:\nolinkurl{10.1145/1458082.1458205}}


\bibitem[Ma et~al\mbox{.}(2018)]%
        {ma2018modeling}
\bibfield{author}{\bibinfo{person}{Jiaqi Ma}, \bibinfo{person}{Zhe Zhao}, \bibinfo{person}{Xinyang Yi}, \bibinfo{person}{Jilin Chen}, \bibinfo{person}{Lichan Hong}, {and} \bibinfo{person}{Ed~H Chi}.} \bibinfo{year}{2018}\natexlab{}.
\newblock \showarticletitle{Modeling task relationships in multi-task learning with multi-gate mixture-of-experts}. In \bibinfo{booktitle}{\emph{Proceedings of the 24th ACM SIGKDD international conference on knowledge discovery \& data mining}}. \bibinfo{pages}{1930--1939}.
\newblock


\bibitem[Milogradskii et~al\mbox{.}(2024)]%
        {milogradskii2024revisiting}
\bibfield{author}{\bibinfo{person}{Aleksandr Milogradskii}, \bibinfo{person}{Oleg Lashinin}, \bibinfo{person}{Alexander P}, \bibinfo{person}{Marina Ananyeva}, {and} \bibinfo{person}{Sergey Kolesnikov}.} \bibinfo{year}{2024}\natexlab{}.
\newblock \showarticletitle{Revisiting BPR: A Replicability Study of a Common Recommender System Baseline}. In \bibinfo{booktitle}{\emph{Proceedings of the 18th ACM Conference on Recommender Systems}}. \bibinfo{pages}{267--277}.
\newblock


\bibitem[Mnih and Salakhutdinov(2007)]%
        {mnih2007pmf}
\bibfield{author}{\bibinfo{person}{Andriy Mnih} {and} \bibinfo{person}{Russ~R Salakhutdinov}.} \bibinfo{year}{2007}\natexlab{}.
\newblock \showarticletitle{Probabilistic Matrix Factorization}. In \bibinfo{booktitle}{\emph{Advances in Neural Information Processing Systems}}, \bibfield{editor}{\bibinfo{person}{J.~Platt}, \bibinfo{person}{D.~Koller}, \bibinfo{person}{Y.~Singer}, {and} \bibinfo{person}{S.~Roweis}} (Eds.), Vol.~\bibinfo{volume}{20}. \bibinfo{publisher}{Curran Associates, Inc.}
\newblock
\urldef\tempurl%
\url{https://proceedings.neurips.cc/paper_files/paper/2007/file/d7322ed717dedf1eb4e6e52a37ea7bcd-Paper.pdf}
\showURL{%
\tempurl}


\bibitem[Nasir and Ezeife(2023)]%
        {nasir2023survey}
\bibfield{author}{\bibinfo{person}{Mahreen Nasir} {and} \bibinfo{person}{Christie~I Ezeife}.} \bibinfo{year}{2023}\natexlab{}.
\newblock \showarticletitle{A survey and taxonomy of sequential recommender systems for e-commerce product recommendation}.
\newblock \bibinfo{journal}{\emph{SN Computer Science}} \bibinfo{volume}{4}, \bibinfo{number}{6} (\bibinfo{year}{2023}), \bibinfo{pages}{708}.
\newblock


\bibitem[Petrov and Macdonald(2022a)]%
        {petrov2022recencysampling}
\bibfield{author}{\bibinfo{person}{Aleksandr Petrov} {and} \bibinfo{person}{Craig Macdonald}.} \bibinfo{year}{2022}\natexlab{a}.
\newblock \showarticletitle{Effective and Efficient Training for Sequential Recommendation using Recency Sampling}. In \bibinfo{booktitle}{\emph{Sixteen ACM Conference on Recommender Systems}}.
\newblock


\bibitem[Petrov and Macdonald(2022b)]%
        {petrov2022systematic}
\bibfield{author}{\bibinfo{person}{Aleksandr Petrov} {and} \bibinfo{person}{Craig Macdonald}.} \bibinfo{year}{2022}\natexlab{b}.
\newblock \showarticletitle{A systematic review and replicability study of bert4rec for sequential recommendation}. In \bibinfo{booktitle}{\emph{Proceedings of the 16th ACM Conference on Recommender Systems}}. \bibinfo{pages}{436--447}.
\newblock


\bibitem[Petrov and Macdonald(2023)]%
        {petrov2023gsasrec}
\bibfield{author}{\bibinfo{person}{Aleksandr~Vladimirovich Petrov} {and} \bibinfo{person}{Craig Macdonald}.} \bibinfo{year}{2023}\natexlab{}.
\newblock \showarticletitle{gsasrec: Reducing overconfidence in sequential recommendation trained with negative sampling}. In \bibinfo{booktitle}{\emph{Proceedings of the 17th ACM Conference on Recommender Systems}}. \bibinfo{pages}{116--128}.
\newblock


\bibitem[Petrov and Macdonald(2025)]%
        {petrov2025improving}
\bibfield{author}{\bibinfo{person}{Aleksandr~Vladimirovich Petrov} {and} \bibinfo{person}{Craig Macdonald}.} \bibinfo{year}{2025}\natexlab{}.
\newblock \showarticletitle{Improving Effectiveness by Reducing Overconfidence in Large Catalogue Sequential Recommendation with gBCE loss}.
\newblock \bibinfo{journal}{\emph{ACM Transactions on Recommender Systems}} \bibinfo{volume}{3}, \bibinfo{number}{4} (\bibinfo{year}{2025}), \bibinfo{pages}{1--34}.
\newblock


\bibitem[Petrov et~al\mbox{.}(2025)]%
        {petrov2025efficient}
\bibfield{author}{\bibinfo{person}{Aleksandr~V Petrov}, \bibinfo{person}{Craig Macdonald}, {and} \bibinfo{person}{Nicola Tonellotto}.} \bibinfo{year}{2025}\natexlab{}.
\newblock \showarticletitle{Efficient Recommendation with Millions of Items by Dynamic Pruning of Sub-Item Embeddings}.
\newblock \bibinfo{journal}{\emph{arXiv preprint arXiv:2505.00560}} (\bibinfo{year}{2025}).
\newblock


\bibitem[Quadrana et~al\mbox{.}(2017)]%
        {quadrana2017personalizing}
\bibfield{author}{\bibinfo{person}{Massimo Quadrana}, \bibinfo{person}{Alexandros Karatzoglou}, \bibinfo{person}{Bal{\'a}zs Hidasi}, {and} \bibinfo{person}{Paolo Cremonesi}.} \bibinfo{year}{2017}\natexlab{}.
\newblock \showarticletitle{Personalizing session-based recommendations with hierarchical recurrent neural networks}. In \bibinfo{booktitle}{\emph{proceedings of the Eleventh ACM Conference on Recommender Systems}}. \bibinfo{pages}{130--137}.
\newblock


\bibitem[Ren et~al\mbox{.}(2019)]%
        {ren2019repeatnet}
\bibfield{author}{\bibinfo{person}{Pengjie Ren}, \bibinfo{person}{Zhumin Chen}, \bibinfo{person}{Jing Li}, \bibinfo{person}{Zhaochun Ren}, \bibinfo{person}{Jun Ma}, {and} \bibinfo{person}{Maarten De~Rijke}.} \bibinfo{year}{2019}\natexlab{}.
\newblock \showarticletitle{Repeatnet: A repeat aware neural recommendation machine for session-based recommendation}. In \bibinfo{booktitle}{\emph{Proceedings of the AAAI conference on artificial intelligence}}, Vol.~\bibinfo{volume}{33}. \bibinfo{pages}{4806--4813}.
\newblock


\bibitem[Ren et~al\mbox{.}(2024)]%
        {ren2024representation}
\bibfield{author}{\bibinfo{person}{Xubin Ren}, \bibinfo{person}{Wei Wei}, \bibinfo{person}{Lianghao Xia}, \bibinfo{person}{Lixin Su}, \bibinfo{person}{Suqi Cheng}, \bibinfo{person}{Junfeng Wang}, \bibinfo{person}{Dawei Yin}, {and} \bibinfo{person}{Chao Huang}.} \bibinfo{year}{2024}\natexlab{}.
\newblock \showarticletitle{Representation learning with large language models for recommendation}. In \bibinfo{booktitle}{\emph{Proceedings of the ACM Web Conference 2024}}. \bibinfo{pages}{3464--3475}.
\newblock


\bibitem[Rendle et~al\mbox{.}(2009)]%
        {rendle2009bpr}
\bibfield{author}{\bibinfo{person}{Steffen Rendle}, \bibinfo{person}{Christoph Freudenthaler}, \bibinfo{person}{Zeno Gantner}, {and} \bibinfo{person}{Lars Schmidt-Thieme}.} \bibinfo{year}{2009}\natexlab{}.
\newblock \showarticletitle{BPR: Bayesian personalized ranking from implicit feedback}. In \bibinfo{booktitle}{\emph{Proceedings of the Twenty-Fifth Conference on Uncertainty in Artificial Intelligence}} (Montreal, Quebec, Canada) \emph{(\bibinfo{series}{UAI '09})}. \bibinfo{publisher}{AUAI Press}, \bibinfo{address}{Arlington, Virginia, USA}, \bibinfo{pages}{452–461}.
\newblock
\showISBNx{9780974903958}


\bibitem[Rendle et~al\mbox{.}(2010)]%
        {rendle2010factorizing}
\bibfield{author}{\bibinfo{person}{Steffen Rendle}, \bibinfo{person}{Christoph Freudenthaler}, {and} \bibinfo{person}{Lars Schmidt-Thieme}.} \bibinfo{year}{2010}\natexlab{}.
\newblock \showarticletitle{Factorizing personalized markov chains for next-basket recommendation}. In \bibinfo{booktitle}{\emph{Proceedings of the 19th international conference on World wide web}}. \bibinfo{pages}{811--820}.
\newblock


\bibitem[Shazeer et~al\mbox{.}(2017)]%
        {shazeer2017outrageously}
\bibfield{author}{\bibinfo{person}{Noam Shazeer}, \bibinfo{person}{Azalia Mirhoseini}, \bibinfo{person}{Krzysztof Maziarz}, \bibinfo{person}{Andy Davis}, \bibinfo{person}{Quoc Le}, \bibinfo{person}{Geoffrey Hinton}, {and} \bibinfo{person}{Jeff Dean}.} \bibinfo{year}{2017}\natexlab{}.
\newblock \showarticletitle{Outrageously Large Neural Networks: The Sparsely-Gated Mixture-of-Experts Layer}. In \bibinfo{booktitle}{\emph{International Conference on Learning Representations}}.
\newblock


\bibitem[Shehzad and Jannach(2023)]%
        {shehzad2023everyone}
\bibfield{author}{\bibinfo{person}{Faisal Shehzad} {and} \bibinfo{person}{Dietmar Jannach}.} \bibinfo{year}{2023}\natexlab{}.
\newblock \showarticletitle{Everyone’s a winner! on hyperparameter tuning of recommendation models}. In \bibinfo{booktitle}{\emph{Proceedings of the 17th ACM Conference on Recommender Systems}}. \bibinfo{pages}{652--657}.
\newblock


\bibitem[Sun et~al\mbox{.}(2019)]%
        {sun2019bert4rec}
\bibfield{author}{\bibinfo{person}{Fei Sun}, \bibinfo{person}{Jun Liu}, \bibinfo{person}{Jian Wu}, \bibinfo{person}{Changhua Pei}, \bibinfo{person}{Xiao Lin}, \bibinfo{person}{Wenwu Ou}, {and} \bibinfo{person}{Peng Jiang}.} \bibinfo{year}{2019}\natexlab{}.
\newblock \showarticletitle{BERT4Rec: Sequential recommendation with bidirectional encoder representations from transformer}. In \bibinfo{booktitle}{\emph{Proceedings of the 28th ACM international conference on information and knowledge management}}. \bibinfo{pages}{1441--1450}.
\newblock


\bibitem[Tang and Wang(2018)]%
        {tang2018personalized}
\bibfield{author}{\bibinfo{person}{Jiaxi Tang} {and} \bibinfo{person}{Ke Wang}.} \bibinfo{year}{2018}\natexlab{}.
\newblock \showarticletitle{Personalized top-n sequential recommendation via convolutional sequence embedding}. In \bibinfo{booktitle}{\emph{Proceedings of the eleventh ACM international conference on web search and data mining}}. \bibinfo{pages}{565--573}.
\newblock


\bibitem[Tanjim et~al\mbox{.}(2020)]%
        {tanjim2020attentive}
\bibfield{author}{\bibinfo{person}{Md~Mehrab Tanjim}, \bibinfo{person}{Congzhe Su}, \bibinfo{person}{Ethan Benjamin}, \bibinfo{person}{Diane Hu}, \bibinfo{person}{Liangjie Hong}, {and} \bibinfo{person}{Julian McAuley}.} \bibinfo{year}{2020}\natexlab{}.
\newblock \showarticletitle{Attentive Sequential Models of Latent Intent for Next Item Recommendation}. In \bibinfo{booktitle}{\emph{Proceedings of The Web Conference 2020}} (Taipei, Taiwan) \emph{(\bibinfo{series}{WWW '20})}. \bibinfo{publisher}{Association for Computing Machinery}, \bibinfo{address}{New York, NY, USA}, \bibinfo{pages}{2528–2534}.
\newblock
\showISBNx{9781450370233}


\bibitem[Tay et~al\mbox{.}(2018)]%
        {tay2018multi}
\bibfield{author}{\bibinfo{person}{Yi Tay}, \bibinfo{person}{Anh~Tuan Luu}, {and} \bibinfo{person}{Siu~Cheung Hui}.} \bibinfo{year}{2018}\natexlab{}.
\newblock \showarticletitle{Multi-pointer co-attention networks for recommendation}. In \bibinfo{booktitle}{\emph{Proceedings of the 24th ACM SIGKDD international conference on knowledge discovery \& data mining}}. \bibinfo{pages}{2309--2318}.
\newblock


\bibitem[Vaswani et~al\mbox{.}(2017)]%
        {vaswani2017attention}
\bibfield{author}{\bibinfo{person}{Ashish Vaswani}, \bibinfo{person}{Noam Shazeer}, \bibinfo{person}{Niki Parmar}, \bibinfo{person}{Jakob Uszkoreit}, \bibinfo{person}{Llion Jones}, \bibinfo{person}{Aidan~N Gomez}, \bibinfo{person}{\L~ukasz Kaiser}, {and} \bibinfo{person}{Illia Polosukhin}.} \bibinfo{year}{2017}\natexlab{}.
\newblock \showarticletitle{Attention is All you Need}. In \bibinfo{booktitle}{\emph{Advances in Neural Information Processing Systems}}, \bibfield{editor}{\bibinfo{person}{I.~Guyon}, \bibinfo{person}{U.~Von Luxburg}, \bibinfo{person}{S.~Bengio}, \bibinfo{person}{H.~Wallach}, \bibinfo{person}{R.~Fergus}, \bibinfo{person}{S.~Vishwanathan}, {and} \bibinfo{person}{R.~Garnett}} (Eds.), Vol.~\bibinfo{volume}{30}. \bibinfo{publisher}{Curran Associates, Inc.}
\newblock
\urldef\tempurl%
\url{https://proceedings.neurips.cc/paper_files/paper/2017/file/3f5ee243547dee91fbd053c1c4a845aa-Paper.pdf}
\showURL{%
\tempurl}


\bibitem[Wang et~al\mbox{.}(2023)]%
        {wang2023sequential}
\bibfield{author}{\bibinfo{person}{Chenyang Wang}, \bibinfo{person}{Weizhi Ma}, \bibinfo{person}{Chong Chen}, \bibinfo{person}{Min Zhang}, \bibinfo{person}{Yiqun Liu}, {and} \bibinfo{person}{Shaoping Ma}.} \bibinfo{year}{2023}\natexlab{}.
\newblock \showarticletitle{Sequential Recommendation with Multiple Contrast Signals}.
\newblock \bibinfo{journal}{\emph{ACM Trans. Inf. Syst.}} \bibinfo{volume}{41}, \bibinfo{number}{1}, Article \bibinfo{articleno}{11} (\bibinfo{date}{Jan.} \bibinfo{year}{2023}), \bibinfo{numpages}{27}~pages.
\newblock
\showISSN{1046-8188}
\href{https://doi.org/10.1145/3522673}{doi:\nolinkurl{10.1145/3522673}}


\bibitem[Wang et~al\mbox{.}(2013)]%
        {wang2013collaborative}
\bibfield{author}{\bibinfo{person}{Hao Wang}, \bibinfo{person}{Binyi Chen}, {and} \bibinfo{person}{Wu-Jun Li}.} \bibinfo{year}{2013}\natexlab{}.
\newblock \showarticletitle{Collaborative topic regression with social regularization for tag recommendation.}. In \bibinfo{booktitle}{\emph{IJCAI}}, Vol.~\bibinfo{volume}{13}. \bibinfo{pages}{2719--2725}.
\newblock


\bibitem[Wang et~al\mbox{.}(2020)]%
        {wang2020next}
\bibfield{author}{\bibinfo{person}{Jianling Wang}, \bibinfo{person}{Kaize Ding}, \bibinfo{person}{Liangjie Hong}, \bibinfo{person}{Huan Liu}, {and} \bibinfo{person}{James Caverlee}.} \bibinfo{year}{2020}\natexlab{}.
\newblock \showarticletitle{Next-item Recommendation with Sequential Hypergraphs}. In \bibinfo{booktitle}{\emph{Proceedings of the 43rd International ACM SIGIR Conference on Research and Development in Information Retrieval}} (Virtual Event, China) \emph{(\bibinfo{series}{SIGIR '20})}. \bibinfo{publisher}{Association for Computing Machinery}, \bibinfo{address}{New York, NY, USA}, \bibinfo{pages}{1101–1110}.
\newblock
\showISBNx{9781450380164}
\href{https://doi.org/10.1145/3397271.3401133}{doi:\nolinkurl{10.1145/3397271.3401133}}


\bibitem[Wang et~al\mbox{.}(2021)]%
        {wang2021session}
\bibfield{author}{\bibinfo{person}{Jianling Wang}, \bibinfo{person}{Kaize Ding}, \bibinfo{person}{Ziwei Zhu}, {and} \bibinfo{person}{James Caverlee}.} \bibinfo{year}{2021}\natexlab{}.
\newblock \showarticletitle{Session-based recommendation with hypergraph attention networks}. In \bibinfo{booktitle}{\emph{Proceedings of the 2021 SIAM international conference on data mining (SDM)}}. SIAM, \bibinfo{pages}{82--90}.
\newblock


\bibitem[Wang et~al\mbox{.}(2015)]%
        {wang2015learning}
\bibfield{author}{\bibinfo{person}{Pengfei Wang}, \bibinfo{person}{Jiafeng Guo}, \bibinfo{person}{Yanyan Lan}, \bibinfo{person}{Jun Xu}, \bibinfo{person}{Shengxian Wan}, {and} \bibinfo{person}{Xueqi Cheng}.} \bibinfo{year}{2015}\natexlab{}.
\newblock \showarticletitle{Learning Hierarchical Representation Model for NextBasket Recommendation}. In \bibinfo{booktitle}{\emph{Proceedings of the 38th International ACM SIGIR Conference on Research and Development in Information Retrieval}} (Santiago, Chile) \emph{(\bibinfo{series}{SIGIR '15})}. \bibinfo{publisher}{Association for Computing Machinery}, \bibinfo{address}{New York, NY, USA}, \bibinfo{pages}{403–412}.
\newblock
\showISBNx{9781450336215}
\href{https://doi.org/10.1145/2766462.2767694}{doi:\nolinkurl{10.1145/2766462.2767694}}


\bibitem[Wolf et~al\mbox{.}(2020)]%
        {wolf-etal-2020-transformers}
\bibfield{author}{\bibinfo{person}{Thomas Wolf}, \bibinfo{person}{Lysandre Debut}, \bibinfo{person}{Victor Sanh}, \bibinfo{person}{Julien Chaumond}, \bibinfo{person}{Clement Delangue}, \bibinfo{person}{Anthony Moi}, \bibinfo{person}{Pierric Cistac}, \bibinfo{person}{Tim Rault}, \bibinfo{person}{Rémi Louf}, \bibinfo{person}{Morgan Funtowicz}, \bibinfo{person}{Joe Davison}, \bibinfo{person}{Sam Shleifer}, \bibinfo{person}{Patrick von Platen}, \bibinfo{person}{Clara Ma}, \bibinfo{person}{Yacine Jernite}, \bibinfo{person}{Julien Plu}, \bibinfo{person}{Canwen Xu}, \bibinfo{person}{Teven~Le Scao}, \bibinfo{person}{Sylvain Gugger}, \bibinfo{person}{Mariama Drame}, \bibinfo{person}{Quentin Lhoest}, {and} \bibinfo{person}{Alexander~M. Rush}.} \bibinfo{year}{2020}\natexlab{}.
\newblock \showarticletitle{Transformers: State-of-the-Art Natural Language Processing}. In \bibinfo{booktitle}{\emph{Proceedings of the 2020 Conference on Empirical Methods in Natural Language Processing: System Demonstrations}}. \bibinfo{publisher}{Association for Computational Linguistics}, \bibinfo{address}{Online}, \bibinfo{pages}{38--45}.
\newblock
\urldef\tempurl%
\url{https://www.aclweb.org/anthology/2020.emnlp-demos.6}
\showURL{%
\tempurl}


\bibitem[Xu et~al\mbox{.}(2024b)]%
        {xu2024MoME}
\bibfield{author}{\bibinfo{person}{Jiahui Xu}, \bibinfo{person}{Lu Sun}, {and} \bibinfo{person}{Dengji Zhao}.} \bibinfo{year}{2024}\natexlab{b}.
\newblock \showarticletitle{MoME: Mixture-of-Masked-Experts for Efficient Multi-Task Recommendation}. In \bibinfo{booktitle}{\emph{Proceedings of the 47th International ACM SIGIR Conference on Research and Development in Information Retrieval}} (Washington DC, USA) \emph{(\bibinfo{series}{SIGIR '24})}. \bibinfo{publisher}{Association for Computing Machinery}, \bibinfo{address}{New York, NY, USA}, \bibinfo{pages}{2527–2531}.
\newblock
\showISBNx{9798400704314}
\href{https://doi.org/10.1145/3626772.3657922}{doi:\nolinkurl{10.1145/3626772.3657922}}


\bibitem[Xu et~al\mbox{.}(2024a)]%
        {xu2024openp5}
\bibfield{author}{\bibinfo{person}{Shuyuan Xu}, \bibinfo{person}{Wenyue Hua}, {and} \bibinfo{person}{Yongfeng Zhang}.} \bibinfo{year}{2024}\natexlab{a}.
\newblock \showarticletitle{Openp5: An open-source platform for developing, training, and evaluating llm-based recommender systems}. In \bibinfo{booktitle}{\emph{Proceedings of the 47th International ACM SIGIR Conference on Research and Development in Information Retrieval}}. \bibinfo{pages}{386--394}.
\newblock


\bibitem[Yan et~al\mbox{.}(2019)]%
        {yan2019cosrec}
\bibfield{author}{\bibinfo{person}{An Yan}, \bibinfo{person}{Shuo Cheng}, \bibinfo{person}{Wang-Cheng Kang}, \bibinfo{person}{Mengting Wan}, {and} \bibinfo{person}{Julian McAuley}.} \bibinfo{year}{2019}\natexlab{}.
\newblock \showarticletitle{CosRec: 2D convolutional neural networks for sequential recommendation}. In \bibinfo{booktitle}{\emph{Proceedings of the 28th ACM international conference on information and knowledge management}}. \bibinfo{pages}{2173--2176}.
\newblock


\bibitem[Yang et~al\mbox{.}(2022)]%
        {yang2022multi}
\bibfield{author}{\bibinfo{person}{Yuhao Yang}, \bibinfo{person}{Chao Huang}, \bibinfo{person}{Lianghao Xia}, \bibinfo{person}{Yuxuan Liang}, \bibinfo{person}{Yanwei Yu}, {and} \bibinfo{person}{Chenliang Li}.} \bibinfo{year}{2022}\natexlab{}.
\newblock \showarticletitle{Multi-Behavior Hypergraph-Enhanced Transformer for Sequential Recommendation}. In \bibinfo{booktitle}{\emph{Proceedings of the 28th ACM SIGKDD Conference on Knowledge Discovery and Data Mining}} (Washington DC, USA) \emph{(\bibinfo{series}{KDD '22})}. \bibinfo{publisher}{Association for Computing Machinery}, \bibinfo{address}{New York, NY, USA}, \bibinfo{pages}{2263–2274}.
\newblock
\showISBNx{9781450393850}
\href{https://doi.org/10.1145/3534678.3539342}{doi:\nolinkurl{10.1145/3534678.3539342}}


\bibitem[Ying et~al\mbox{.}(2018)]%
        {ying2018sequential}
\bibfield{author}{\bibinfo{person}{Haochao Ying}, \bibinfo{person}{Fuzhen Zhuang}, \bibinfo{person}{Fuzheng Zhang}, \bibinfo{person}{Yanchi Liu}, \bibinfo{person}{Guandong Xu}, \bibinfo{person}{Xing Xie}, \bibinfo{person}{Hui Xiong}, {and} \bibinfo{person}{Jian Wu}.} \bibinfo{year}{2018}\natexlab{}.
\newblock \showarticletitle{Sequential recommender system based on hierarchical attention network}. In \bibinfo{booktitle}{\emph{IJCAI international joint conference on artificial intelligence}}.
\newblock


\bibitem[Yuan et~al\mbox{.}(2019)]%
        {yuan2019asimple}
\bibfield{author}{\bibinfo{person}{Fajie Yuan}, \bibinfo{person}{Alexandros Karatzoglou}, \bibinfo{person}{Ioannis Arapakis}, \bibinfo{person}{Joemon~M. Jose}, {and} \bibinfo{person}{Xiangnan He}.} \bibinfo{year}{2019}\natexlab{}.
\newblock \showarticletitle{A Simple Convolutional Generative Network for Next Item Recommendation}. In \bibinfo{booktitle}{\emph{Proceedings of the Twelfth ACM International Conference on Web Search and Data Mining}} (Melbourne VIC, Australia) \emph{(\bibinfo{series}{WSDM '19})}. \bibinfo{publisher}{Association for Computing Machinery}, \bibinfo{address}{New York, NY, USA}, \bibinfo{pages}{582–590}.
\newblock
\showISBNx{9781450359405}
\href{https://doi.org/10.1145/3289600.3290975}{doi:\nolinkurl{10.1145/3289600.3290975}}


\bibitem[Yuan et~al\mbox{.}(2023)]%
        {yuan2023go}
\bibfield{author}{\bibinfo{person}{Zheng Yuan}, \bibinfo{person}{Fajie Yuan}, \bibinfo{person}{Yu Song}, \bibinfo{person}{Youhua Li}, \bibinfo{person}{Junchen Fu}, \bibinfo{person}{Fei Yang}, \bibinfo{person}{Yunzhu Pan}, {and} \bibinfo{person}{Yongxin Ni}.} \bibinfo{year}{2023}\natexlab{}.
\newblock \showarticletitle{Where to go next for recommender systems? id-vs. modality-based recommender models revisited}. In \bibinfo{booktitle}{\emph{Proceedings of the 46th International ACM SIGIR Conference on Research and Development in Information Retrieval}}. \bibinfo{pages}{2639--2649}.
\newblock


\bibitem[Zheng et~al\mbox{.}(2017)]%
        {zheng2017joint}
\bibfield{author}{\bibinfo{person}{Lei Zheng}, \bibinfo{person}{Vahid Noroozi}, {and} \bibinfo{person}{Philip~S. Yu}.} \bibinfo{year}{2017}\natexlab{}.
\newblock \showarticletitle{Joint Deep Modeling of Users and Items Using Reviews for Recommendation}. In \bibinfo{booktitle}{\emph{Proceedings of the Tenth ACM International Conference on Web Search and Data Mining}} (Cambridge, United Kingdom) \emph{(\bibinfo{series}{WSDM '17})}. \bibinfo{publisher}{Association for Computing Machinery}, \bibinfo{address}{New York, NY, USA}, \bibinfo{pages}{425–434}.
\newblock
\showISBNx{9781450346757}
\href{https://doi.org/10.1145/3018661.3018665}{doi:\nolinkurl{10.1145/3018661.3018665}}


\bibitem[Zhou et~al\mbox{.}(2022)]%
        {zhou2022mixture}
\bibfield{author}{\bibinfo{person}{Yanqi Zhou}, \bibinfo{person}{Tao Lei}, \bibinfo{person}{Hanxiao Liu}, \bibinfo{person}{Nan Du}, \bibinfo{person}{Yanping Huang}, \bibinfo{person}{Vincent Zhao}, \bibinfo{person}{Andrew~M Dai}, \bibinfo{person}{Quoc~V Le}, \bibinfo{person}{James Laudon}, {et~al\mbox{.}}} \bibinfo{year}{2022}\natexlab{}.
\newblock \showarticletitle{Mixture-of-experts with expert choice routing}.
\newblock \bibinfo{journal}{\emph{Advances in Neural Information Processing Systems}}  \bibinfo{volume}{35} (\bibinfo{year}{2022}), \bibinfo{pages}{7103--7114}.
\newblock


\end{thebibliography}


\end{document}